\documentclass[fleqn,usenatbib]{mnras}

\usepackage[T1]{fontenc}
\usepackage{ae,aecompl}

\usepackage{graphicx}	
\usepackage{amsmath}	
\usepackage{amssymb}	
\usepackage{bm}
\usepackage{tabularx} 
\usepackage{pdflscape}
\usepackage{color}
\usepackage{times}
\usepackage{hyperref}

\makeatletter
\newcommand{\HI}{{\rm H\,{\scriptstyle I}}}
\newcommand{\HII}{{\rm H\,{\scriptstyle II}}}

\newcommand{\HeII}{{\rm He\,{\scriptstyle II}}}
\newcommand{\HeIII}{{\rm He\,{\scriptstyle III}}}
\newcommand{\CIV}{{\rm C\,{\scriptstyle IV}}}

\newcommand{\NV}{{\rm N\,{\scriptstyle V}}}

\newcommand{\OI}{{\rm O\,{\scriptstyle I}}}

\newcommand{\Rmnum}[1]{\expandafter\@slowromancap\romannumeral #1@}

\newcommand{\xHI}{x_{\mbox{\tiny H\Rmnum{1}}}}

\newcommand{\NHI}{N_{\mbox{\tiny H\Rmnum{1}}}}

\newcommand{\CDDF}{\frac{\partial^2\mathcal{N}}{\partial\NHI\partial z}}
\newcommand{\sigmaHI}{\sigma_{\mbox{\tiny H\Rmnum{1}}}}
\newcommand{\Muv}{M_{\mbox{\tiny UV}}}
\newcommand{\br}{\boldsymbol{r}}

\newcommand{\ignore}[1]{}

\makeatother

\title[What reionised the Universe?]
{The Role of Galaxies and AGN in Reionising the IGM - \Rmnum{1}: Keck Spectroscopy of $\boldsymbol 5$ < $\boldsymbol z$ < $\boldsymbol 7$ Galaxies in the QSO Field J1148+5251}
\author[K. Kakiichi et al.] 
{Koki Kakiichi$^1$\thanks{E-mail: k.kakiichi@ucl.ac.uk}, Richard  S. Ellis$^1$, Nicolas Laporte$^1$, Adi Zitrin$^2$, Anna-Christina Eilers$^3$, \newauthor  
Emma Ryan-Weber$^{4,5}$, Romain A. Meyer$^1$, Brant Robertson$^6$, Daniel P. Stark$^7$ and  \newauthor Sarah E. I.  Bosman$^1$ \\
$^1$ Department of Physics and Astronomy, University College London, London, WC1E 6BT, UK \\
$^2$ Physics Department, Ben-Gurion University of the Negev, PO Box 653, Beer-Sheva 84105, Israel \\
$^3$ Max Planck Institute for Astronomy, K\"onigstuhl 17, 69117 Heidelberg, Germany \\
$^4$ Centre for Astrophysics and Supercomputing, Swinburne University of Technology, Hawthorn, VIC 3122, Australia \\
$^5$ ARC Centre of Excellence for All Sky Astrophysics in 3 Dimensions (ASTRO 3D), Australia \\
$^6$ Department of Astronomy and Astrophysics, University of California, Santa Cruz, 1156 High Street, Santa Cruz, CA 95064 USA \\
$^7$ Steward Observatory, University of Arizona, Tucson AZ 85719 USA \\
}
\begin{document}
\label{firstpage}
\pagerange{\pageref{firstpage}--\pageref{lastpage}} \pubyear{2018}
\maketitle


\begin{abstract}
We introduce a new method for determining the influence of galaxies and active galactic nuclei (AGN) on the intergalactic medium (IGM) at high redshift and illustrate its potential via a first application to the field of the $z=6.42$ QSO J1148+5251. Correlating spatial positions Lyman break galaxies (LBGs) with the Lyman alpha forest seen in the spectrum of a background QSO, we provide a statistical measure of the typical escape fraction of Lyman continuum photons. Using Keck DEIMOS spectroscopy to locate 7 colour-selected LBGs in the range $5.3\lesssim z\lesssim 6.4$ we examine the spatial correlation between this sample and Ly$\alpha$/Ly$\beta$ transmission fluctuations in a Keck ESI spectrum of the QSO. Interpreting the statistical $\HI$ proximity effect as arising from faint galaxies clustered around the LBGs, we translate the observed mean Ly$\alpha$ transmitted flux into a constraint on the mean escape fraction $\langle f_{\rm esc}\rangle\geq0.08$ at $z\simeq6$. We also report individual transverse $\HI$ proximity effect for a $z=6.177$ luminous LBG via a Ly$\beta$ transmission spike and two broad Ly$\alpha$ transmission spikes around the $z=5.701$ AGN. We discuss the origin of such associations which suggest that while faint galaxies are primarily driving reionisation, luminous galaxies and AGN may provide important contributions to the UV background or thermal fluctuations of the IGM at $z\simeq6$. Although a limited sample, our results demonstrate the potential of making progress using this method in resolving one of the most challenging aspects of the contribution of galaxies and AGN to cosmic reionisation.
\end{abstract}


\section{Introduction}
Understanding how and when cosmic reionisation occurred represents one of the most important challenges in observational cosmology and galaxy formation. Of particular interest is the nature of sources responsible, which was first discussed over 50 years ago \citep{1965ApJ...142.1633G}. Although reionisation is commonly assumed to be driven by the abundant population of intrinsically faint star-forming galaxies (e.g. \citealt{Robertson2013,Robertson2015}, for a review see \citealt{Stark2016}), a key assumption is that the average escape fraction of Lyman continuum (LyC) photons is $\sim10-20\rm~\%$. Such high escape fractions are rarely encountered in lower redshift star-forming galaxies where direct measurements of the LyC leakage is possible \citep{Mostardi2015,Naidu2017}. On the other hand, recent observations of Ly$\alpha$ emission in the spectra of $z>7$ galaxies \citep{Oesch2015,Zitrin2015} might indicate that reionisation is accelerated in the volumes around the most luminous galaxies \citep{Stark2017}, possibly as a result of their harboring active galactic nuclei (AGN, \citealt{Laporte2017}). A significant contribution of ionising photons from rare sources such as luminous galaxies and/or AGN (\citealt{Giallongo2015}, but see \citealt{Parsa2018}) may also explain the significant scatter in the effective optical depth of Ly$\alpha$ absorption in the spectra of $z\gtrsim5.5$ QSOs \citep{Becker2015a,Chardin2015,Chardin2017,Bosman2018}. However, both observationally and theoretically the relative ionising contribution of galaxies and AGN is a subject of intense debate \citep{Madau2015,DAloisio2017,Qin2017,Hassan2018,Mitra2018}.

A fundamental impasse to progress is the absence of a reliable technique to measure the escape fraction $f_{\rm esc}$ of ionising photons at high redshift where direct measures of the leaking LyC radiation become impractical due to foreground line-of-sight absorption. Indirect methods have been examined including absorption line measures of the covering fraction of low ionisation gas in the spectra of lensed galaxies \citep{Jones2013,Leethochawalit2016} which suggest a modest increase in $f_{\rm esc}$ to $z\simeq4$, but the method assumes low ionisation gas is a faithful tracer, geometrically and kinematically, of neutral hydrogen \citep{Reddy2016,Vasei2016}. Other methods such as the analysis of recombination lines \citep{Zackrisson2013,Zackrisson2017}, requires access to Balmer lines seen beyond 2 microns at high redshift and also necessitates an accurate knowledge of the nature of the stellar population. 

In this paper we propose a new method for estimating $f_{\rm esc}$ at high redshift which is based on examining the cross-correlation between star-forming galaxies and the Ly$\alpha$ absorption spectrum of a background QSO probed in the same cosmic volume. Such an approach \citep{Adelberger2003,Adelberger2005} has been productive at $z\simeq2-3$ in exploring associations between galaxies and QSOs and their immediate environments \citep{Rudie2012,Prochaska2013,Turner2014}, as well as in studies of the reionisation of $\HeII$ \citep{Schmidt2017}. However, the idea is largely unexploited in the $\HI$ reionisation era other than studies by \citet{Diaz11,Diaz2014,Diaz2015} (also \citealt{2017MNRAS.469L..53G,Cai2017}) which focused on the environs of $\CIV$ absorption systems at z$\sim$5.7. In this first paper in the series, we develop the method which exploits the statistical association between star-forming galaxies proximate to the QSO sightline and fluctuations in the Ly$\alpha$ forest in the QSO spectrum. We illustrate the potential via an application to a cosmic volume spanning the redshift range $5.3\lesssim z\lesssim 6.4$ in the field of the $z$=6.42 SDSS QSO J1148+5251.

To test the influence of star-forming galaxies and AGN on reionisation we propose to establish {\it a direct connection between the distribution of galaxies of known redshift and luminosity and the physical state of the IGM in the same cosmic volume}. In this paper we introduce the methodology of how the population-averaged LyC escape fraction can thus be determined. High resolution spectroscopy of a $z>6$ QSO provides the redshift-dependent Ly$\alpha$ forest transmission of the IGM and the photoionisation rate $\Gamma_{\rm HI}$ of the UV background, with the aid of cosmological simulations \citep{Bolton2007,FG2008,Becker2013}. Additionally, spectroscopic follow up of colour-selected Lyman break galaxies (LBGs) provides UV luminosities $L_{\rm UV}$ and precise redshifts in the same volume probed by the background QSO. By predicting the number of ionising photons emitted from survey galaxies, we can evaluate the contribution of galaxies to the observationally-measured UV background. Using the spectroscopically-detected luminous LBGs as signposts (e.g. using the host-halo mass) it is possible to estimate the abundance of (unseen) fainter galaxies clustered around them.  For this we utilise the results of deeper imaging data which has established the galaxy-halo connection from joint analyses of the well-established luminosity function down to $M_{\rm UV}\simeq-15$ \citep{Bouwens2015,Bouwens2017,Atek2015,Livermore2017,Ishigaki2018,Ono2018} and clustering measurements \citep{McLure2009,Barone-Nugent2014,Harikane2016,Harikane2018} in the context of  $\Lambda$CDM cosmology \citep[e.g.][]{Springel2005,vandenBosch2013}. The population-averaged LyC escape fraction, $\langle f_{\rm esc}\rangle$ is then obtained by equating the total ionising output from the combined population of luminous and fainter galaxies with the photoionisation rate of the IGM.


A plan of the paper follows. In Section~\ref{sec:observations} we introduce the necessary observations for our programme which includes broad-band photometry necessary for colour-selection of $z>5$ Lyman break galaxies, Keck spectroscopy using the wide-field DEIMOS spectrograph which yields precise redshifts essential for accurate mapping, and the archival ESI spectrum of QSO J1148+5251. We use these data to produce a catalog of star-forming galaxies as well as the Ly$\alpha$ transmission spectrum in the same redshift range. We analyse our observations in Section~\ref{sec:result}, calculating the correlation between our spectroscopically-confirmed galaxies and the fluctuations in the Ly$\alpha$ forest which gives us the mean Ly$\alpha$ transmitted flux around galaxies. In Section~\ref{sec:interpretation} we discuss the physical origin of the observed Ly$\alpha$ transmitted flux around LBGs and introduce our methodology which takes into account the associated but fainter galaxies which are undetected in our imaging survey thereby deriving a mean escape fraction of LyC photons at $z\simeq$6. The result is presented in Section~\ref{sec:fesc}.  In Section~\ref{sec:ind} we examine two specific cases where sources can be directly associated with features in the Ly$\alpha$ forest which provides insight into the possible contribution of rarer, luminous sources including AGN. In Section~\ref{sec:discussion} we discuss the promise and challenges of our new method and the prospects with further data. 

Throughout this paper we adopt the Planck 2015 cosmology ($\Omega_m, \Omega_\Lambda, \Omega_b, h, \sigma_8, n_s$)=(0.3089, 0.6911, 0.04860, 0.6774, 0.8159, 0.9667) \citep{2016A&A...594A..13P}.
We use pkpc and pMpc (ckpc and cMpc) to indicate distances in proper (comoving) units. All magnitudes in this paper are quoted in the AB system \citep{Oke-Gunn1983}.

\section{Observations}\label{sec:observations}

\begin{figure*}
\centering
\includegraphics[width=\columnwidth]{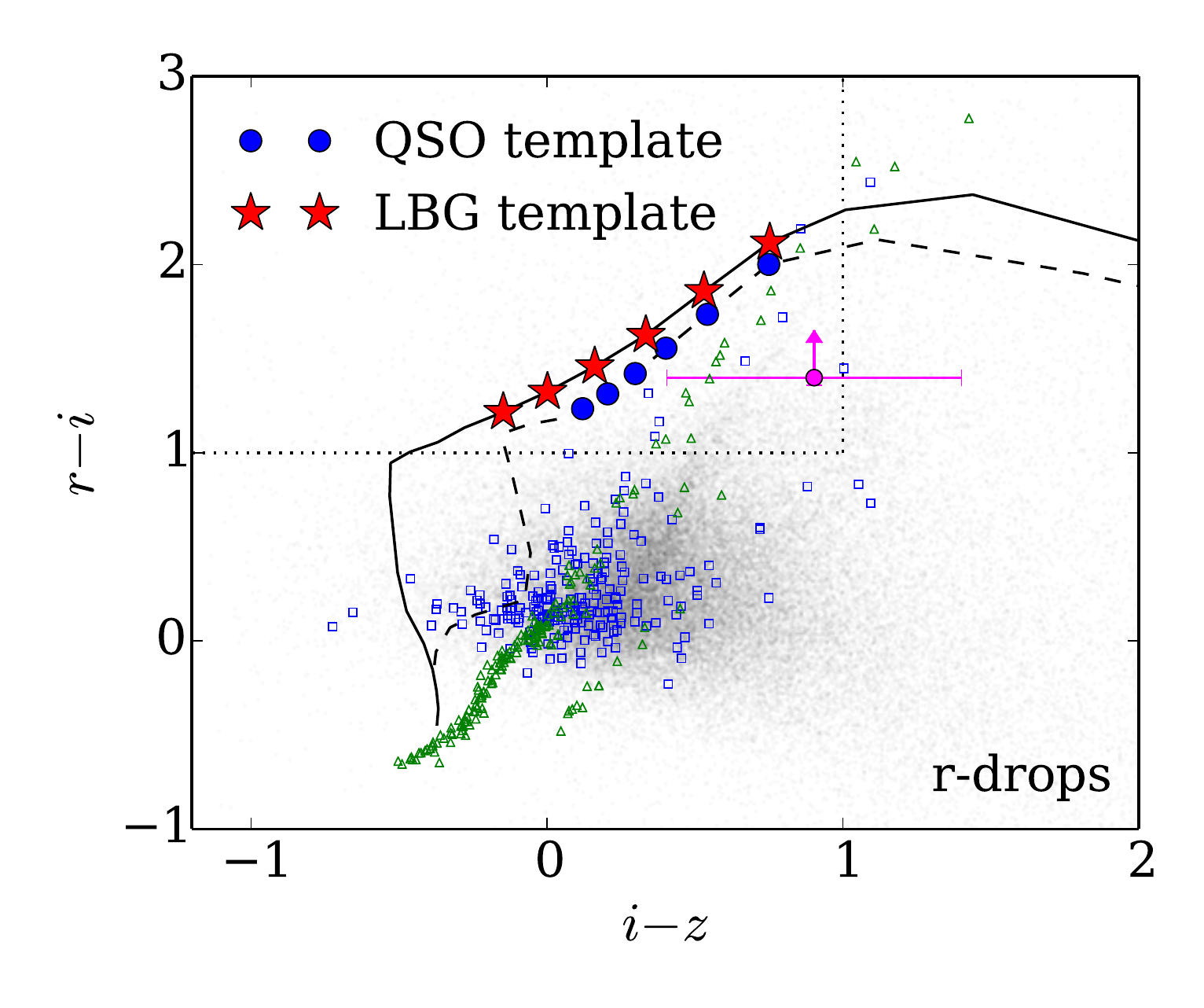}
\includegraphics[width=\columnwidth]{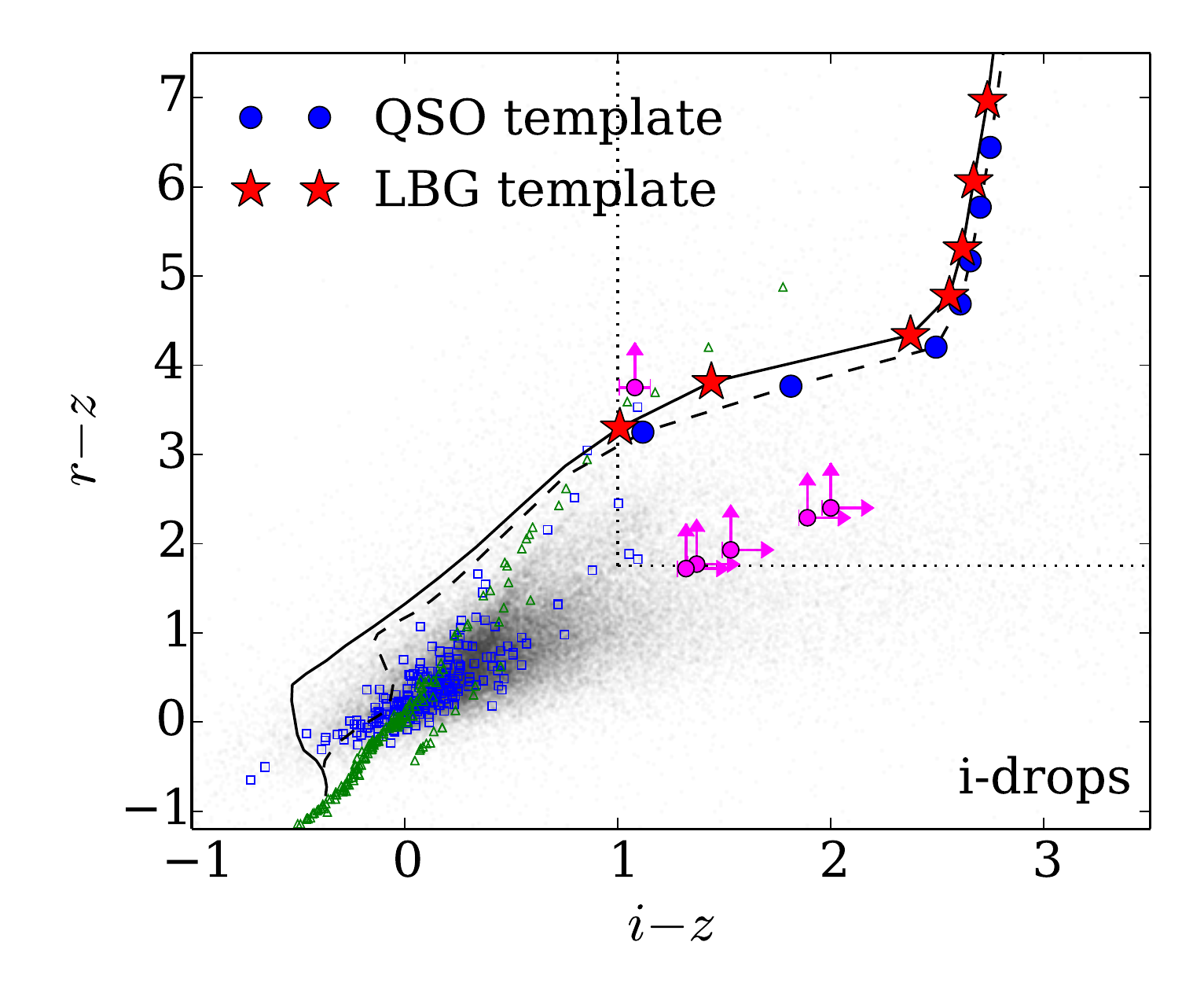}
\caption{Colour-colour diagram for $r$- (left) and $i$-dropouts (right). The locii of LBGs with $W_{\rm Ly\alpha}=50$~\AA~(red star symbols) and a QSO (filled blue symbols) template spectrum from $z=5.2, \dots, 5.7$  (left) and $5.8, \dots, 6.4$ (right) by 0.1 interval are shown. The magenta points are the spectroscopically-confirmed $r,i$-dropouts in the Q1148 field. The small black points represent candidates from the photometric catalogue identified by SExtractor. Typical colours for $0<z<3$ interlopers (open blue squares) from VUDS-DR1 samples in COSMOS field \citep{2015A&A...576A..79L,2017A&A...600A.110T} and for Galactic stars (open green triangles) \citep{1983ApJS...52..121G} are overlaid. Our adopted selection criteria for dropout candidates are indicated by dotted lines.}
\label{fig:color-diagram}
\end{figure*}

Our choice of the SDSS QSO J1148+5251 at $z=6.4189$ (RA= 11h 48m 16.7s +52 deg 51m 50.39s, J2000) for the illustration of our new method was based on the availability of its ESI high signal to noise spectrum and deep ground and space-based imaging from which we can photometrically-select galaxies in the relevant redshift range. For this QSO the uncontaminated Ly$\alpha$ forest spans the redshift range $5.26<z<6.42$.  Archival data from the {\it Spitzer} and {\it Chandra} Space Telescopes provides additional information on the stellar mass and AGN activity of selected sources in the QSO field (e.g. \citealt{Jiang2006,Gallerani2017}). 

\subsection{Imaging data and photometric catalogue}

Deep archival Large Binocular Telescope (LBT) images of the Q1148 field in the SDSS $r$-, $i$-, and $z$-band filters taken by the Large Binocular Camera (LBC) were used to construct a photometric catalogue of $r$- and $i$-dropout candidates for Keck spectroscopic follow-up. LBC pipeline-reduced images reported by \citet{2014A&A...568A...1M} (PI: R. Gilli)\footnote{\url{http://www.oabo.inaf.it/~LBTz6/}} were downloaded from the LBT archive The exposure times were $\sim3\rm~hrs$ in $r$ and $\sim1.5\rm~hrs$ in $i$ and $z$. This panoramic dataset covers a field of $23 \times 25$ arcmin ($\sim39.5\times42.5$ $h^{-1}\rm cMpc$ at $z = 6$) which covers a substantial fraction of the expected mean free path of ionising photons at this epoch, $\lambda_{\rm mfp}\simeq6.0[(1+z)/7]^{-5.4}\rm~pMpc$ \citep{Worseck2014} or $17\rm~arcmin$ in radius. From the processed data, we constructed our own photometric source catalogue using SExtractor  \citep{1996A&AS..117..393B}. The limiting magnitudes in each bandpass were estimated by randomly placing fixed 2 arsec apertures in blank regions. We derived $5\sigma$ limiting magnitudes of $r=26.3$, $i=25.9$, and $z=25.0$ (and at $2\sigma$, $r=27.3$, $i=26.9$, and $z=26.0$) in agreement with the values reported by \citet{2014A&A...568A...1M}. 

\begin{figure}
\hspace*{-0.5cm}
\includegraphics[width=1.1\columnwidth]{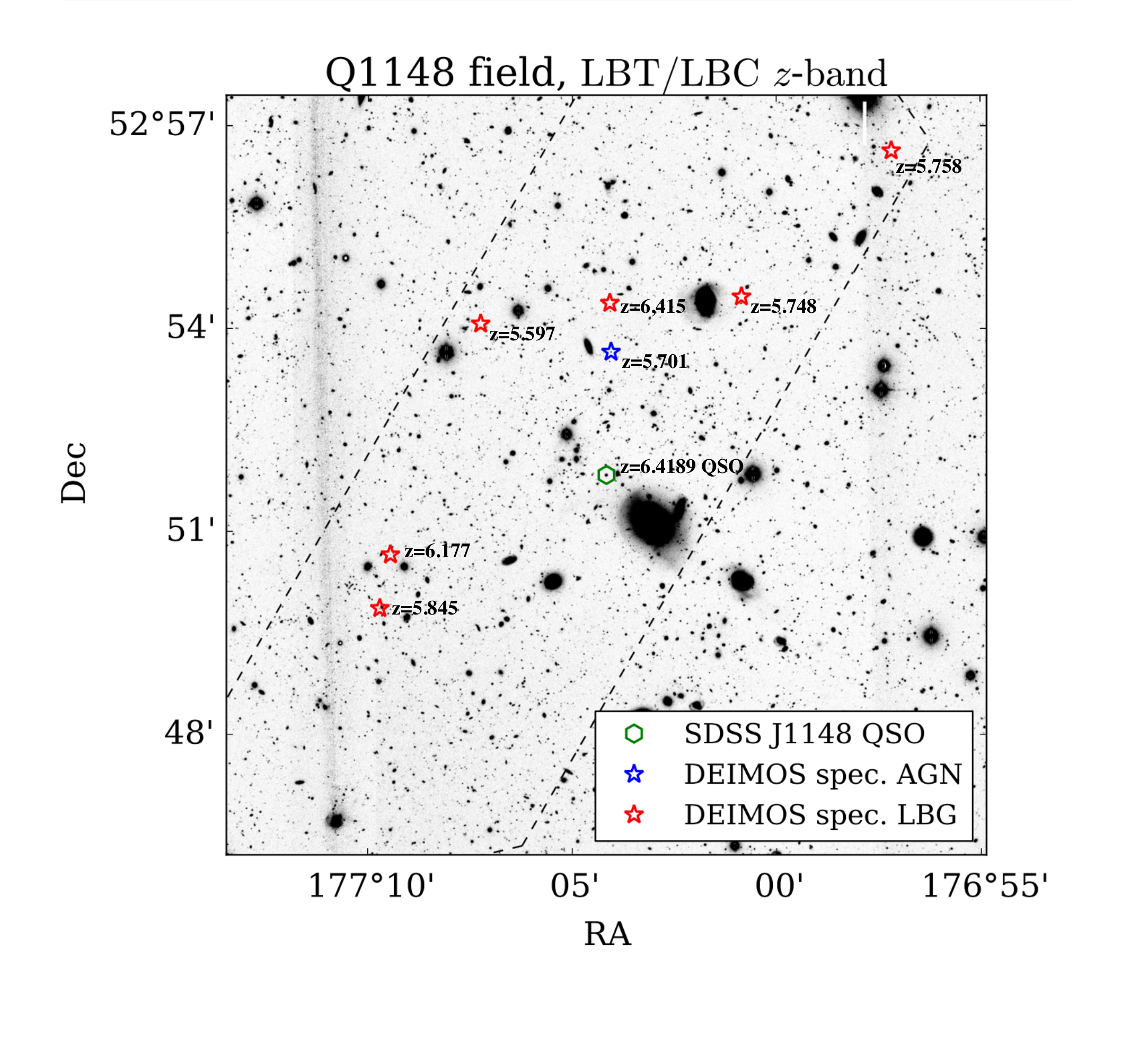}
\vspace{-0.8cm}
\caption{LBT/LBC $z$-image of the Q1148 field overlaid with spectroscopically identified dropouts (star symbols, red: LBG, blue: AGN) and the background SDSS J1148 QSO (diamond). Each symbol is annotated with the spectroscopic redshift. The DEIMOS footprint is marked (dashed).}\label{fig:image}
\label{fig:DEIMOS}
\end{figure}


In order to select our candidate LBGs in the desired redshift range, we imposed a $5\sigma$ detection limit of $z=25.0$ for our primary selection with fainter secondary candidates at the $3\sigma$ limit of $z=25.6$. We selected candidate LBGs in the sought-after redshift range $5.26\lesssim z\lesssim6.42$ according to the following criteria:

\begin{equation}
r-i>1.0~~\mbox{AND}~~i-z<1.0\label{eq:1}
\end{equation}
for $r$-dropouts and 
\begin{equation}
i-z>1.0~~\mbox{AND}~~[r>r(2\sigma)~\mbox{OR}~r-z>1.75]\label{eq:2}
\end{equation}
for $i$-dropouts.

We can visualize the $i$-dropout criteria by considering template spectra for target LBGs and AGN in Figure~\ref{fig:color-diagram}. Here a strong Ly$\alpha$ emission line could produce bluer $i-z$ colours and thus a traditional $i-z>1.3$ colour cut \citep[e.g.][]{2006ApJ...653...53B,2007ApJ...670..928B} would miss a substantial fraction of objects at $5.3<z<5.7$ and $\sim20-30$ \% at $z>5.7$ \citep{2005ApJ...626..666M,Diaz11}. Likewise Type \Rmnum{2} QSOs could have a very blue $i-z<0$ colour at $z>5.3$ due to the strong Ly$\alpha$ emission line \citep{2006MNRAS.365..833M,Diaz11}. In Figure~\ref{fig:color-diagram}, we consider both $r$- and $i$-dropout criteria in the context of the locus of a BPASS galaxy model \citep[version 2.0,][]{BPASS16,BPASS17} with continuous star formation at $100\rm~Myr$ age, $Z=0.20\rm~Z_\odot$ metallicity, and Ly$\alpha$ equivalent width $W_{\rm Ly\alpha}=50~$\AA~, and that of a mean QSO template \citep{Telfer2002} from redshift $5.3$ to $6.4$ at $0.1$ redshift interval in the context of LBT filters\footnote{The filter bandpasses were derived from \url{http://abell.as.arizona.edu/~lbtsci/Instruments/LBC/lbc.html}}.  The IGM transmission is computed using \textsc{IGMtransmission} code \citep{2011arXiv1105.6208H} based on the transmission curves of \citet{2006MNRAS.365..807M}. The adopted selection criteria, Equations (\ref{eq:1}) and ({\ref{eq:2}), are marked. After applying these criteria, two authors (KK and NL) visually inspected all candidates removing sources contaminated with artefacts, diffraction spikes of nearby stars and sources close to the boundaries of the detector mosaic. There are 124 objects in the final photometric catalogue of $r$- and $i$-drop candidates.

\begin{figure}
\centering
\includegraphics[width=0.95\columnwidth]{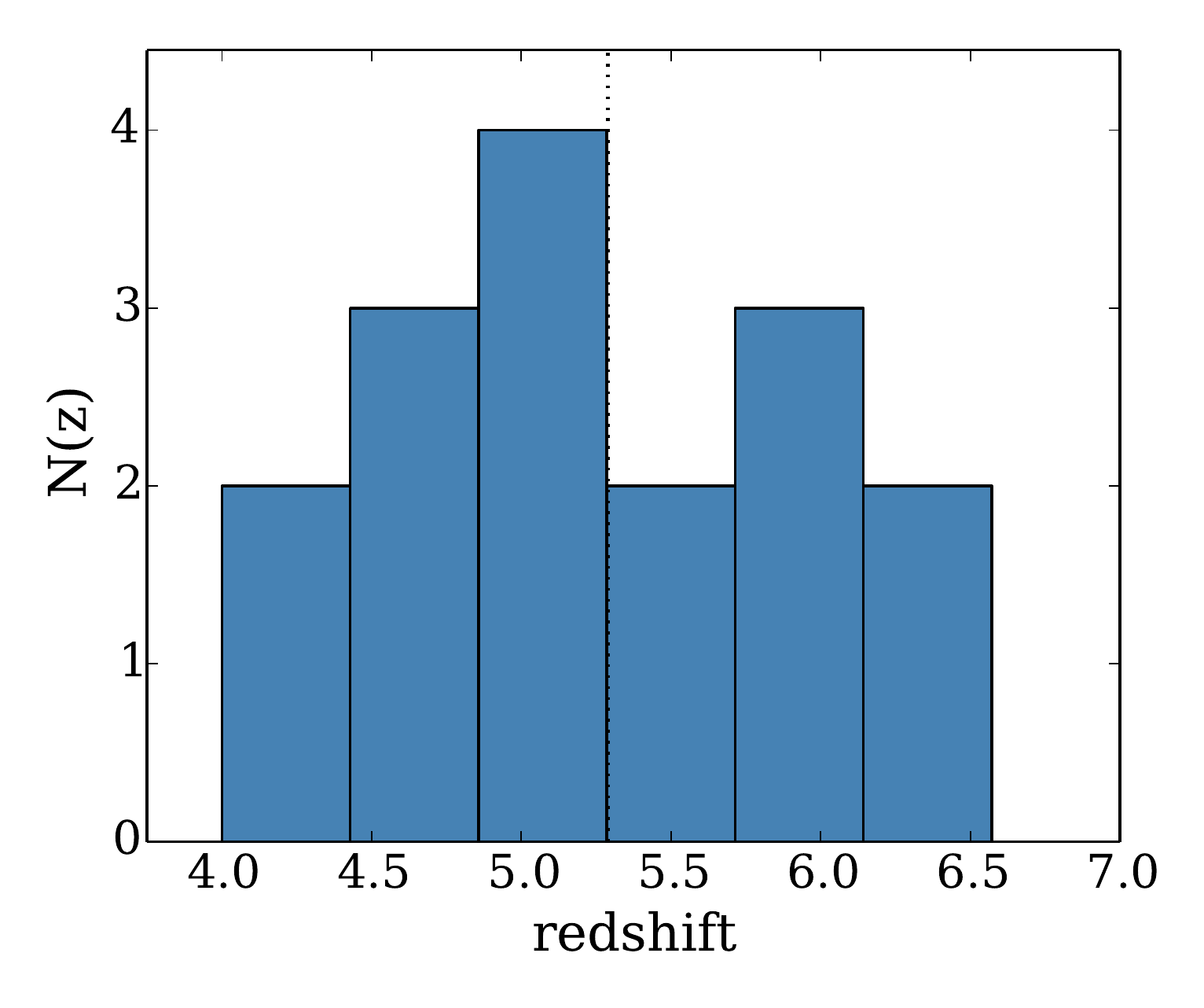}
\vspace{-0.1cm}
\caption{Redshift distribution of the spectroscopically-confirmed LBGs. The dashed line indicates the lower limit at $z$=5.3 for which the Ly$\alpha$ forest can be examined in the ESI spectrum of the QSO.}
\label{fig:N(z)}
\end{figure}

\subsection{Galaxy spectroscopy}

\begin{table*}
\caption{DEIMOS spectroscopic catalog}\label{table:line_inventory}
\begin{tabular}{llllllllll}
\hline\hline
ID  & $z_{\rm Ly\alpha}$$^{a}$  & RA (J2000) & DEC (J2000) & $r~[\rm mag]$ & $i~[\rm mag]$ & $z~[\rm mag]$ & $M_{\rm UV}$$^b$ & Note\\ 
\hline
      002 \ignore{82376}  &  5.701 &  11h48m16.20s &   52d53m39.55s  &  >27.3 & $24.63\pm0.06$ & $23.55\pm0.04$ & $-23.11\pm0.04$ &  AGN (Ly$\alpha$+$\NV$) \\
      004 \ignore{62282} &  6.177 &  11h48m37.80s &   52d50m39.60s	&  >27.3 & >26.9                  & $25.01\pm0.14$ & $-21.78\pm0.14$ & LBG (Ly$\alpha$)\\
      008 \ignore{85398} &  5.597 &  11h48m28.98s &   52d54m04.50s	&  >27.3 & >26.9                  & $25.53\pm0.13$ & $-21.10\pm0.13$  & LBG (Ly$\alpha$)\\
      009 \ignore{87556} &  6.415 &  11h48m16.32s &   52d54m22.94s	&  >27.3 & >26.9                  & $24.90\pm0.12$ & $-21.95\pm0.12$ & LBG (near Q1148) \\
      015 \ignore{57258} &  5.845 &  11h48m38.83s &   52d49m51.97s	&  $26.76\pm0.69$ & $25.36\pm0.39$  & $24.46\pm0.31$ & $-22.24\pm0.31$ & LBG (Ly$\alpha$) \\ 
      022 \ignore{88010} &  5.748 &  11h48m03.42s &   52d54m28.56s	&  >27.3 & >26.9 & $25.37\pm0.15$ & $-21.30\pm0.15$ & LBG (Ly$\alpha$)\\
     043 \ignore{102424} &  5.758 &  11h47m48.72s &   52d56m37.98s	&  >27.3 & >26.9 & $25.58\pm0.16$ & $-21.10\pm0.16$ & LBG (Ly$\alpha$)\\
\hline
\multicolumn{6}{l}{$^a$ By interpreting the peak of the line as Ly$\alpha$ redshift (measured in this work).} \\
\multicolumn{9}{l}{$^b$ Based on the apparent $z$ magnitude, assuming the $k$-correlation $2.5(\alpha-1)\log_{10}(1+z_{\mbox{\tiny Ly$\alpha$}})$ with a spectral slope $\alpha=2$.} 
\end{tabular}
\end{table*}

The photometric candidates were spectroscopically observed through an ongoing survey undertaken with the DEep Imaging Multi-Object Spectrograph (DEIMOS) at the Nasmyth focus of the 10-m Keck \Rmnum{2} telescope \citep{2003SPIE.4841.1657F} on March 26-27 2017 (PI: Zitrin).  Conditions were clear and seeing was typically between 0.9-1.5 arcsec on 26th and 0.7-1.0 arcsec on 27th. We placed one slitmask of $16.7\times5.0$ arcmin$^2$ field of view so as to maximise the number of dropout targets from the LBT photometric catalogue and encompassing a large volume within the mean free path of ionising photons at this epoch (Figure~\ref{fig:image}). In selecting targets for the mask, greater priority was given to $i$-dropouts to increase the likelihood of detecting Ly$\alpha$ emission in redshift range sampled by the Ly$\alpha$ forest, yielding 45 dropout targets in the mask.  A 1.0 arcsec slitwidth was used with the 600 line mm$^{-1}$ grating (600ZD) providing spectroscopic coverage between $4950$~\AA~and $10000$~\AA~with a spectral resolution of $3.5$~\AA\ . The mask was observed for 4.3 hrs. All data were reduced using the \textsc{spec2d} IDL pipeline \citep{2012ascl.soft03003C,2013ApJS..208....5N}. The wavelength calibration was done using the afternoon arc lamp. The final reduction provides two-dimensional (2D) spectra and variance arrays. The spectra were visually inspected for emission lines independently by the four of the authors (KK, RSE, NL, and AZ). Two authors (RSE and NL) were blinded from the locations of transmission features in the QSO spectrum (see below) to avoid unconscious biases. 

In total we secured spectroscopic redshifts for 16 sources including a previously-identified AGN \citep{Mahabal2005}, corresponding to a $\simeq$35 \% success rate of spectroscopic confirmation. All emission lines in each 2D spectrum coincide with the expected location of the dropout target on the slit. The overall redshift distribution of the spectroscopic sample is shown in Figure~\ref{fig:N(z)}. However, due the limited three bands photometry for the Q1148 field, the photometric redshifts were fairly approximate. Within the $5.3<z<6.4$ redshift range which overlaps the volume where the IGM transmission can be traced in the absorption line spectrum of SDSS J1148+5251, we have a sample of 6 spectroscopically-confirmed LBGs plus the AGN (excluding one LBG at $z_{\rm Ly\alpha}$$=$6.415 lying in the proximity zone of the Q1148). Thus the final success rate of finding galaxies in the Ly$\alpha$ forest region was $\simeq$13\%. Spectra of the LBGs and AGN are shown in Figure~\ref{fig:agn} and \ref{fig:spectra}. The properties of the sources in the relevant redshift range for this study are listed in Table~\ref{table:line_inventory}.

\begin{figure}
\hspace{-0.8cm}
\includegraphics[width=1.15\columnwidth]{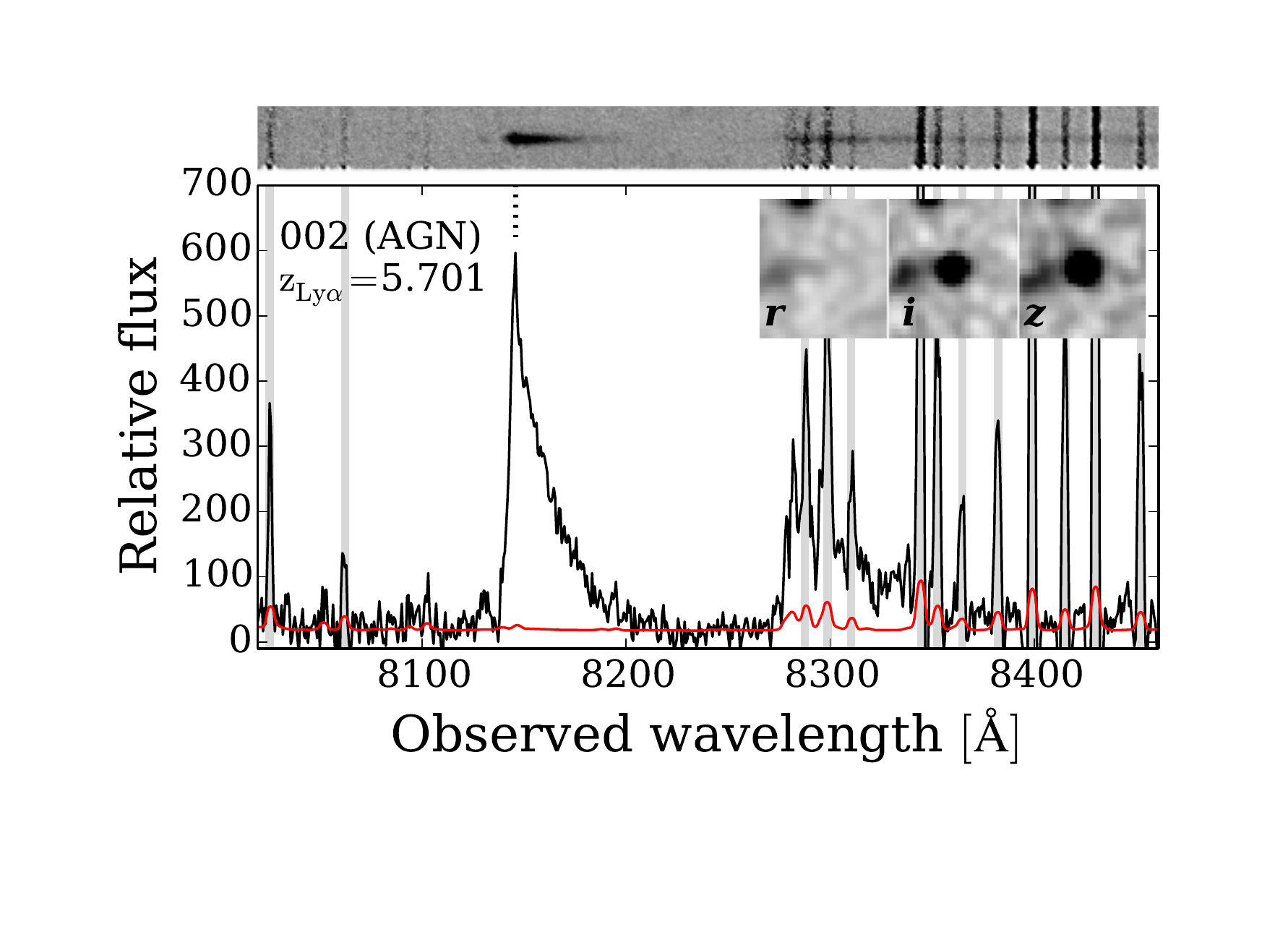}
\vspace{-1.3cm}
\caption{The spectroscopically-confirmed faint AGN at $=5.701$ in the Q1148 field (black: flux, red: noise) with DEIMOS. Skylines \citep{Osterbrock1996} are masked (grey shaded). The 2D spectrum (top panel) and the postage stamp $riz$ image (inset) are shown.  This source (RD J1148+5253) was previously identified by \citet{Mahabal2005}. }
\label{fig:agn}
\end{figure}

\begin{figure*}
\centering
\vspace{0.1cm}
\includegraphics[width=0.85\textwidth]{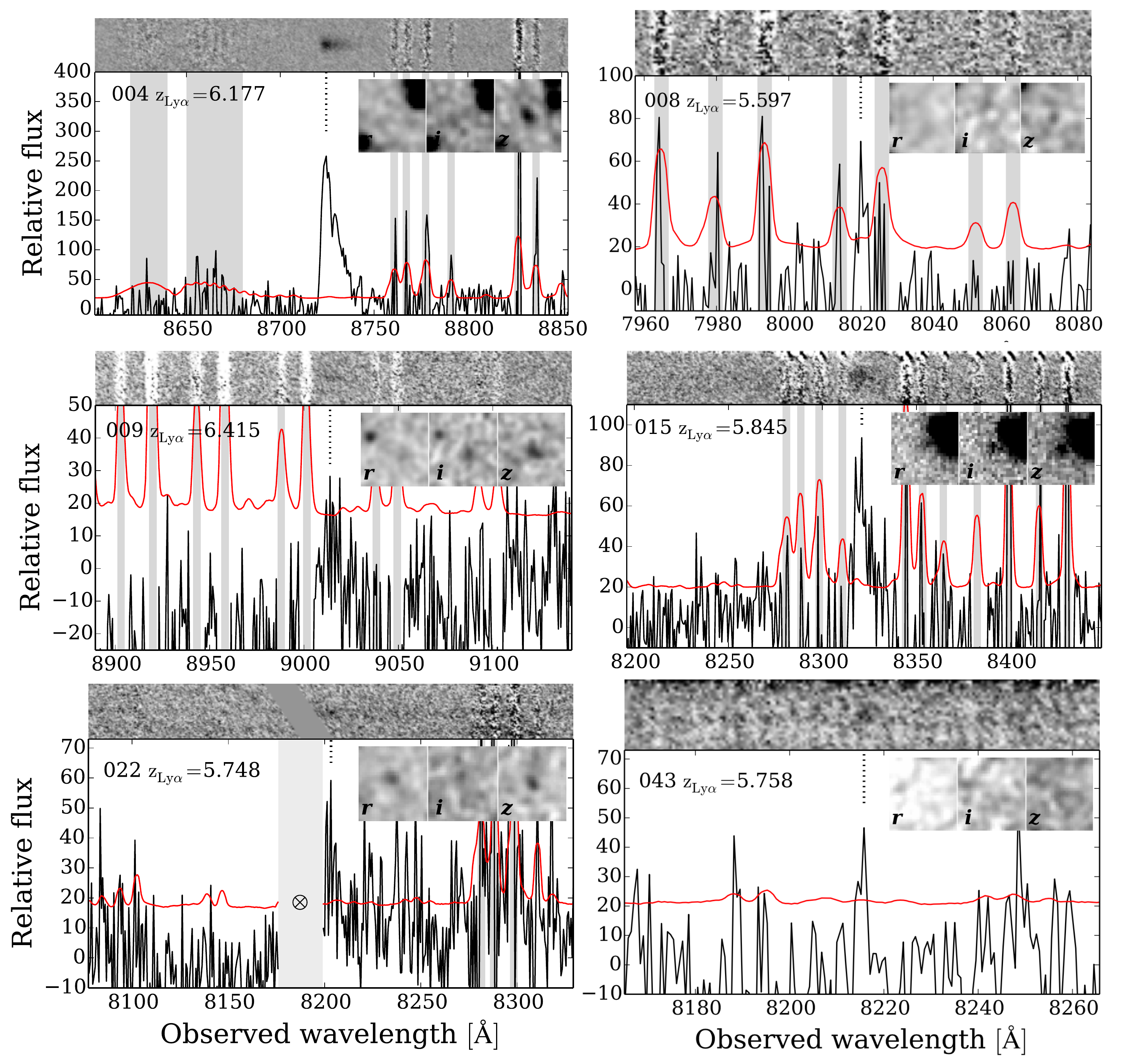}
\vspace{0.1cm}
\caption{Same as Figure~\ref{fig:agn}, but for spectroscopically-confirmed $z>5.3$ LBGs in the Q1148 field.}
\label{fig:spectra}
\end{figure*}

\subsection{QSO spectroscopy \& Ly$\alpha$ transmission features}

To examine the structure in the Ly$\alpha$ forest of SDSS J1148+5251 QSO \citep{Fan2003}, we used a spectrum taken with the Echellette Spectrograph and Imager (ESI) at the Keck \Rmnum{2} telescope from a large sample of QSOs uniformly reduced by \citet{Eilers2017} (Figure \ref{fig:continuum}). The systemic redshift of J1148+5251 is taken from the CO redshift presented in \citet{2010ApJ...714..834C}. The spectral resolution is $R\approx5000$ sampled with $\sim$5 pixels ($\simeq10\rm~km~s^{-1}$ per pixel) within one resolution element.

To estimate the wavelength-dependent continuum level, we use a principal component analysis (PCA) as described by \citet{Eilers2017}. This PCA-based continuum estimate $C_\lambda$ is used to calculate the Ly$\alpha$ transmitted flux $F_\alpha=e^{-\tau_\alpha}$,
\begin{equation}
F_\alpha=f_\lambda/C_\lambda+n_\lambda/C_\lambda,
\end{equation}
where $f_\lambda$ is the observed flux and $n_\lambda$ is the noise in the Q1148 ESI spectrum.

\begin{figure*}
\includegraphics[width=\textwidth]{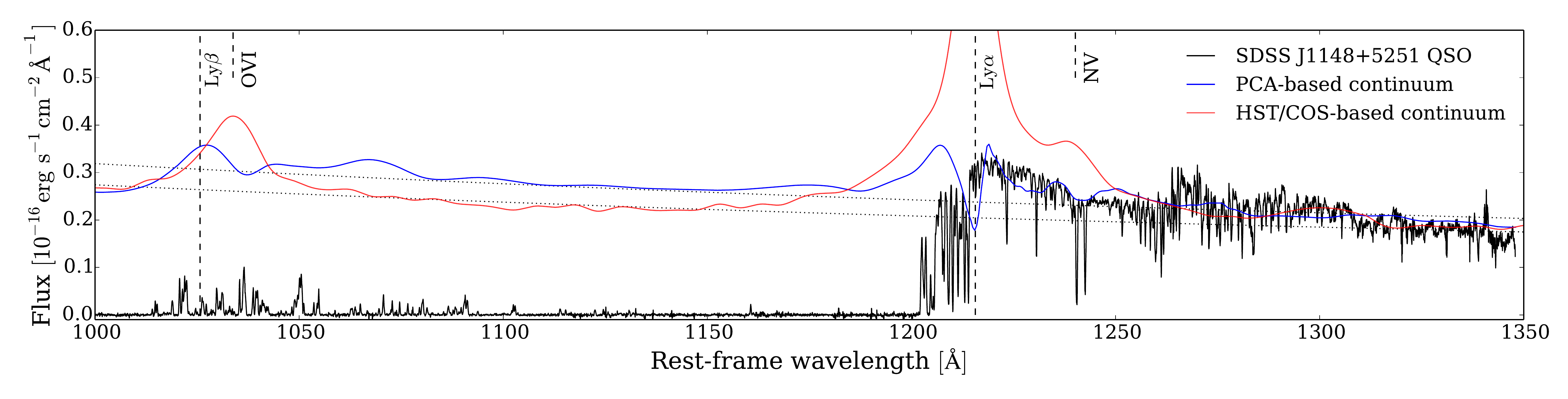}
\vspace{-0.5cm}
\caption{ESI spectrum and estimated continuum level for the SDSS J1148+5251 $z=6.4189$ QSO (black). The latter is based on the PCA spectrum (blue) and HST/COS spectrum (red: continuum \ignore{gray: full spectrum}of SDSS J0929+4644 $z=0.24$ QSO). The dotted lines indicate a power-law continuum with $\alpha_\nu=-0.5$.}
\label{fig:continuum}
\end{figure*}

To estimate the uncertainty, we also employed an empirical technique based on HST/COS spectra of $z\lesssim1$ UV-bright AGN \citep{2016ApJ...817..111D}\footnote{Publicly available online: \url{https://archive.stsci.edu/prepds/igm/}}. The continuum level was then estimated for the subset of 17 HST/COS continuum spectra classified as type `QSO'. We compared the continuum redward of the Ly$\alpha$ emission line of the HST/COS spectra with the Q1148 ESI spectrum and derived the best-fit continuum by minimising the chi-square for $>1270$~\AA~.   Although Q1148 has a weak Ly$\alpha$ emission line, unlike those in the set of the 17 HST/COS spectra, this only affects the derived Ly$\alpha$ absorption properties in the vicinity of the QSO, which is not used in the subsequent analysis. Comparing the Ly$\alpha$ transmitted flux between the PCA-based and HST/COS-based methods, the difference in the continuum level is $\simeq20$ per cent level at median over the redshift range $5.5<z_{\rm Ly\alpha}<6.3$.  This is sufficiently small not to affect the subsequent analysis and results in this paper.

We identify Ly$\alpha$ and Ly$\beta$ transmission spikes using an automated wavelet-based algorithm. We correlate (i.e. wavelet transform) the continuuum-normalized Ly$\alpha$ forest spectrum with a `Mexican hat' wavelet $\psi_\sigma(x)\propto\sigma^{-1/2}(1-(x/\sigma)^2)\exp(-x^2/2\sigma^2)$ (normalised with $\int\psi_\sigma(x)dx=0$),
\begin{equation}
w_\sigma(\lambda)=\int F_\alpha(\lambda)\psi_{\sigma}(\lambda-\lambda')d\lambda'.
\end{equation}
The width of the wavelet was varies according to $\sigma=10,\dots,250\rm~km~s^{-1}$ with $10\rm~km~s^{-1}$ interval. At each wavelength pixel, we record the maximum wavelet coefficient $\displaystyle w_{\rm max}(\lambda)=\max_{\sigma\in{\rm all}}w_\sigma(\lambda)$ for all width choice. Robust transmission spikes are chosen as the local maxima of the wavelet coefficients, $w_{\rm max}(\lambda)$, whose signal-to-noise ratio at a peak pixel is larger than $5\sigma$. The wavelet-based estimate of the widths of the transmission spikes are recorded as the width at which gives the local maxima of the wavelet coefficients. The method successfully identifies the previous known Ly$\alpha$ transmission spike at $z=6.083$ \citep{White2003,White2005,Oh2005}. The list of the identified Ly$\alpha$ and Ly$\beta$ transmission spikes is tabulated in Table~\ref{table:spikes}.

\section{Galaxy-Ly$\alpha$ forest cross-correlations}\label{sec:result}

We now introduce the observed correlation between galaxies and Ly$\alpha$ transmission features in the J1148 QSO field. We focus initially on the 3D mapping of galaxies as it relates to identifiable Ly$\alpha$ transmission spikes and absorption troughs. We then examine the statistical correlation between spectroscopically-confirmed galaxies and the Ly$\alpha$ transmitted flux.  Later, in Section~\ref{sec:interpretation} we discuss the physical basis of this cross-correlation signal and develop a methodology in order to derive a constraint on the mean LyC escape fraction at $z\sim6$ in Section \ref{sec:fesc}.

\subsection{The observed distribution of galaxies around Ly$\alpha$ transmission spikes and absorption troughs}
\label{sec:mapping}

\begin{figure*}
\includegraphics[width=\textwidth]{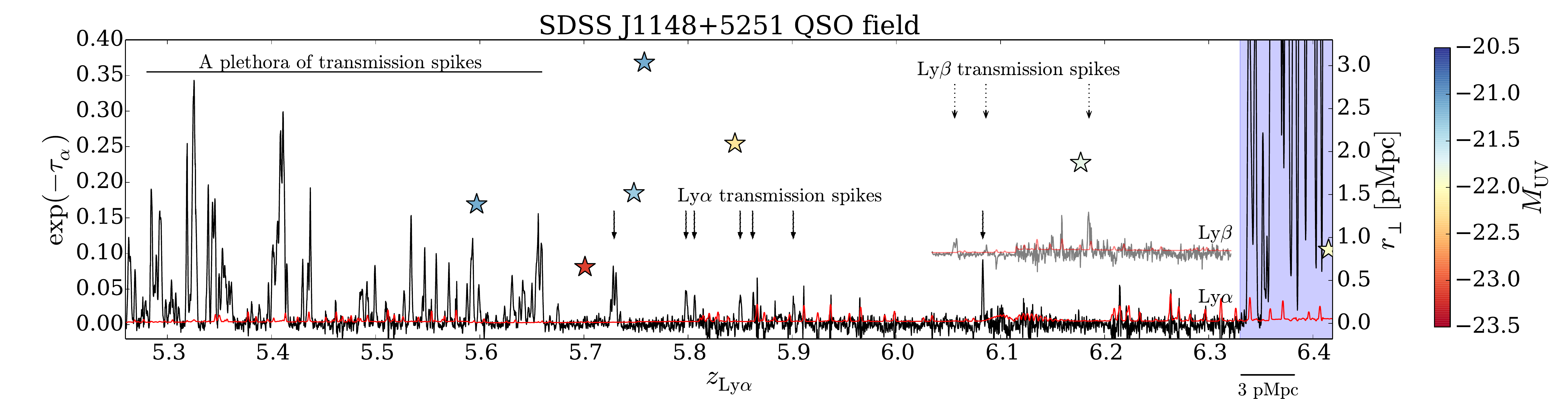}
\vspace{-0.2cm}
\caption{Continuum normalised ESI spectrum (black) of the QSO J1148+5251 plotted in terms of the Ly$\alpha$ opacity,  $\exp(-\tau_\alpha)$, alongside the spectroscopically-confirmed $r,i$-dropout galaxies (star symbols). The noise level in the ESI spectrum is shown in red and incorporates OH and O$_2$ sky line residuals. The region of the absorption spectrum covering the Ly$\beta$ forest (between the Ly$\alpha$ and Ly$\gamma$ forest regions) is shown in grey and offset vertically for convenience. UV luminosities of the spectroscopic sample are indicated by the colour bar and the y axis refers to the angular distance $r_\perp$ of the galaxies from QSO sightline in proper units. The line-of-sight distance corresponding to $\Delta z$$=$0.05 ($\approx$$3$ pMpc) is indicated by the ruler at the bottom right corner. The proximity zone of J1148+5251 is marked by the blue shaded region. Noticeable Ly$\alpha$ transmission spikes at $z$$>$5.7 are marked with arrows, followed by a plethora of transmission spikes at lower redshifts. }
\label{fig:map}
\end{figure*}

In Figure~\ref{fig:map} we show the spatial distribution of spectroscopically-confirmed galaxies from our DEIMOS survey in the context of Ly$\alpha$ forest transmission spikes and absorption troughs in the ESI spectrum of QSO J1148+5251. The continuum normalised QSO spectrum of the transmitted Ly$\alpha$ flux, $e^{-\tau_\alpha}$, is shown with the Ly$\alpha$ redshifts $z_{\rm Ly\alpha}$ and the physical separation $r_\perp$ of the galaxies relative to the QSO sightline. This 3D mapping of galaxies around the varying Ly$\alpha$ transmission gives us our first glimpse of how galaxies influence the physical state of the IGM at the end of reionisation. Three out of our 6 LBGs (at $z_{\rm Ly\alpha}=5.597, 5.845, 6.177$) lie close to the vicinity of Ly$\alpha$ and/or Ly$\beta$ transmission spikes in the QSO spectrum, while 2 LBGs at $z_{\rm Ly\alpha}=5.748, 5.758$ are located close to deep absorption troughs. One of our LBGs at $z_{\rm Ly\alpha}=6.415$ resides within the proximity zone of the J1148+5251 QSO (indicated by the blue shaded region). The source at $z_{\rm Ly\alpha}=5.701$ is a previously known AGN \citep{Mahabal2005} and its location is bracketed by two broad Ly$\alpha$ transmission spikes \citep{Gallerani2008}. 

It is noteworthy that $\simeq$40\% of our spectroscopic sample is found close to Ly$\alpha$ transmission spikes, particularly since the redshift distribution of $r,i$-dropout selection is quite broad \citep{Vanzella2009,Stark2010}. However, there may well be selection effects biasing the visibility of Ly$\alpha$ emission in the galaxy sample, e.g. in wavelength regions unaffected by strong skylines. In order to quantify the relative spatial distribution of LBGs and Ly$\alpha$ absorption more rigorously, it is necessary to adopt a statistical approach.

\begin{table}
\centering
\caption{Transmission Features at $z>5.5$ in the Ly$\alpha$ and Ly$\beta$ forest regions in the Q1148 ESI spectrum.}\label{table:spikes}
\begin{tabular}{llll}
\hline\hline
$z$ & $S/N$ & $z$ & $S/N$  \\ 
\hline
\multicolumn{4}{c}{\underline{Ly$\alpha$ transmission spikes}} \\
5.527 &	6.1 &		5.641 &	20.0 \\
5.534 &	30.8 & 	5.647 &	7.0  \\
5.547 &	18.1 & 	5.650 &	16.0  \\
5.551 &	5.5 &		5.657 &	21.5  \\
5.558 &	20.1 &	5.675 &	7.4  \\
5.570 &	18.5 &	5.729 &	8.4  \\
5.588 &	12.3 & 	5.798 &	10.0  \\
5.593 &	32.3 & 	5.806 &	8.5  \\
5.599 &	15.2 & 	5.850 &	9.4  \\
5.624 &	6.4 & 	5.862 &	8.9  \\
5.631 &	22.6 & 	5.901 &	6.6  \\
5.638 &	12.8 & 	6.083 &	15.1  \\
\multicolumn{4}{c}{\underline{Ly$\beta$ transmission spikes}} \\
6.056 &	6.0 &  & \\
6.086 &	6.6 & &\\ 
6.185 &	8.7 & &\\
\hline
\end{tabular}
\end{table}


\subsection{Statistical $\HI$ proximity effect: \newline the mean Ly$\alpha$ transmitted flux around galaxies}\label{sec:stat}

To examine the cross correlation between the location of spectroscopically-confirmed galaxies and Ly$\alpha$ forest absorption features, we compute the mean Ly$\alpha$ transmitted flux, $\langle\exp(-\tau_\alpha(r))\rangle$, around the spectroscopically confirmed LBGs as a function of physical distance $r$ from a galaxy to Ly$\alpha$ forest pixels,  
\begin{equation}
\left\langle \exp(-\tau_\alpha(r))\right\rangle=\frac{\displaystyle\sum_{i<pair(r)} w_iF_{\alpha,i}}{\displaystyle\sum w_i},\label{eq:mean}
\end{equation}
where $F_{\alpha,i}=e^{-\tau_{\alpha,i}}$ is the Ly$\alpha$ transmitted flux at a physical distance $r_i$ from a galaxy of interest. $w_i$ is the weight for galaxy-Ly$\alpha$ forest flux pair, by which we down-weight noisy pixels as $w_i=1/\sigma^2_{N,i}$. The physical radial distance is computed from $r=\sqrt{r_\perp^2+r_\parallel^2}$ where $r_\perp=\theta D_A(z_{\mbox{\tiny LBG}})$ and $r_\parallel=\int_{z_{\rm pixel}}^{z_{\mbox{\tiny LBG}}}cdz/[(H(z)(1+z)]$ where $z_{\mbox{\tiny LBG}}$ and $z_{\rm pixel}$ are the redshifts of a LBG\footnote{We take a Ly$\alpha$ redshift as a galaxy redshift, $z_{\mbox{\tiny Ly$\alpha$}}=z_{\mbox{\tiny LBG}}$. The velocity offsets of Ly$\alpha$ redshifts relative to the systemic galaxy redshifts vary by $\sim0-500\rm~km~s^{-1}$ \citep[e.g.][and references therein]{Mainali2017}. At a typical velocity offset $\simeq200\rm~km~s^{-1}$ the systematic error in distance is $\simeq300\rm~pkpc$ at $z=5.8$. While for small-scale  applications this involves a correction \citep{Steidel2010,Turner2014}, this has a negligible effect on the large-scale cross-correlation presented in this paper.} and Ly$\alpha$ forest pixel. We did not divide the  Ly$\alpha$ transmitted flux $F_i$ in each pixel by the mean Ly$\alpha$ transmission $e^{-\bar{\tau}_{\rm eff}(z)}$ (to subtract the mean redshift evolution of the IGM, $\bar{\tau}_{\rm eff}(z)$ \citep[e.g.][]{Fan2006,Becker2013a} because, at $z\gtrsim5.75$, the observed Ly$\alpha$ transmitted flux is below the noise level. While equation (\ref{eq:mean}) gives more weight to the Ly$\alpha$ transmission around lower redshift LBGs, it provides the most direct statistical measurement independent of external constraints. A further advantage of this statistical measure is that we need not apply uncertain completeness corrections to our spectroscopic samples. Our procedure provides a measure of the mean $\HI$ gas density around {\it detected} galaxies. This galaxy-centric view contrasts with Ly$\alpha$ forest-centric statistical measures, e.g. the number of galaxies around Ly$\alpha$ transmission spikes, for which completeness corrections in the galaxy sample would be critical. 

In Figure \ref{fig:model} we show the observed mean Ly$\alpha$ transmitted flux around spectroscopically confirmed LBGs with $5.3<z<6.3$ as a function of proper distance in the Q1148 field. We consider $\langle z\rangle \simeq5.8\pm0.2$ as the representative redshift based on the mean redshift of the LBG sample. The maximum distance ($6\rm~pMpc$) is governed by the typical mean free path of ionising photons at $z\approx6$ \citep{Worseck2014}. The error is estimated using the Jackknife resampling based on 5 sub-samples removing one galaxy at a time. As the two innermost bins at $r<1\rm~pMpc$ are based on only one source, we exclude them from the statistical analysis. Although a modest sample, the data presents tentative, intriguing evidence for an increasing Ly$\alpha$ forest transmission closer to the LBGs. This indicates the presence of statisitcal $\HI$ proximity effect at $z\simeq5.8$. The Spearman rank correlation coefficient is $r_s=-0.47$ which corresponds to a `moderate' correlation at a $\approx80-90$ per cent confidence level \citep{2012psa..book.....W}. The correlation is somewhat weaker if the AGN sample is included, degrading the coefficient to $r_s=-0.30$. 

Our sample probes only one sight line, thus any interpretation of the positive signal is affected by both potential systematic errors and small number statistics. The apparent hump at $r\approx 4\rm~pMpc$ is caused by repeatedly selecting the same prominent Ly$\alpha$ transmission spike at $z\approx5.73$. We have tested this by artificially masking between $z$=5.64 and 5.74, where Ly$\alpha$ forest is likely affected by the proximate  $z=5.701$ AGN, and find that the hump is removed. The Jacknife method likely underestimates the error discussed above as the removal of one source near $z\approx5.7$ contributes little to the variance. At this stage we consider the positive correlation between LBGs and Ly$\alpha$ transmission spikes tentative, but sufficient to demonstrate the potential of our method. Although an increased sample size is clearly required, Figure  \ref{fig:model} demonstrates it is possible to probe the gaseous environment of galaxies at the end of reionisation by a spectroscopic survey in $z>6$ QSO fields. 

\begin{figure*}
\includegraphics[width=1.5\columnwidth]{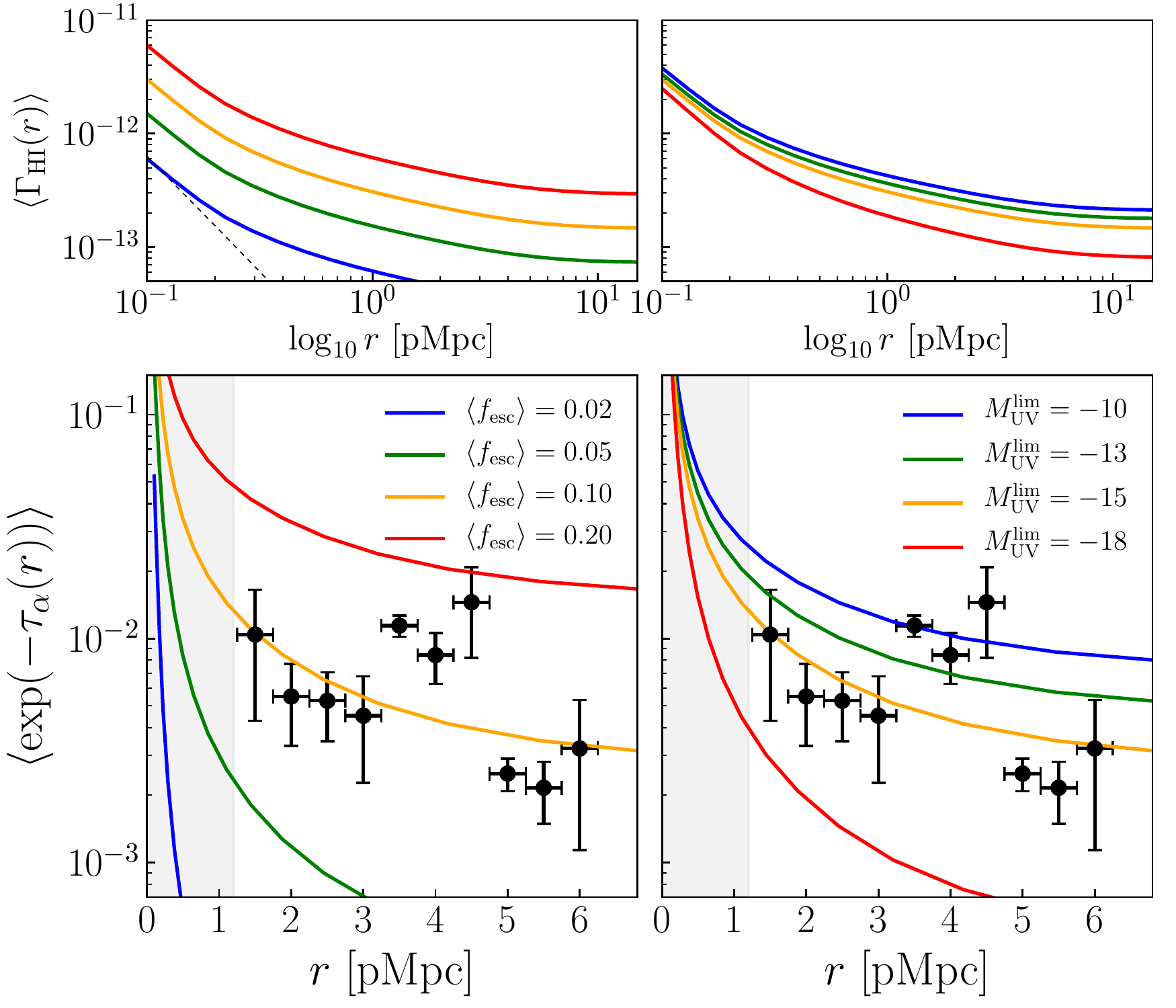}
\caption{Comparison of the observed mean Ly$\alpha$ transmitted flux around LBGs (black) with the theoretical model at $z=5.8$. The model shows the contribution to the photoionisation rate from sub-luminous galaxies clustered around the LBGs for different values of ({\it left panel}) the mean LyC escape fraction and ({\it right panel}) the minimum UV luminosity of ionising galaxies. In the left (right) panel the value of $M_{\rm UV}^{\rm lim}=-15$ ($\langle f_{\rm esc}\rangle=0.10$) is fixed.  The local contribution from a bright LBG alone is indicated as the dotted line. The average photoionisation rate and mean Ly$\alpha$ transmitted flux around LBGs are shown in the top and bottom panels.}\label{fig:model}
\end{figure*}

\section{Interpreting the galaxy-Ly$\alpha$ forest cross-correlations}\label{sec:interpretation}

The $\HI$ proximity effect is normally thought to arise due to the enhanced UV background around ionising sources. In this section we discuss the physical interpretation of the {\it statistical $\HI$ proximity effect} seen in the mean Ly$\alpha$ transmitted flux around LBGs in J1148 QSO field. The basis of our method will be to assume that this statistical $\HI$ proximity effect arises not only from the detected LBGs but also from undetected faint galaxies which cluster around them. By balancing the ionising output of this combined population of luminous and fainter galaxies and the UV background via the statistical $\HI$ proximity effect, we can constrain the population-averaged LyC escape fraction at $z\simeq6$. Although the fainter sources cannot be detected in our observing campaign,  we will use our spectroscopically-detected luminous LBGs effectively as signposts, indicating their likely presence as predicted both by deeper imaging observations and expectations of hierarchical clustering in $\Lambda$CDM cosmology. 

\subsection{Methodology}

In order to interpret our data, we have developed a simple radiative transfer model to examine the influence of galaxies on the IGM. Later we use the model to fit the observed mean Ly$\alpha$ transmitted flux around LBGs to derive a constraint on LyC escape fraction. Although more approximate than one based on numerical radiative transfer or radiation hydrodynamic simulations, it has the benefit of illustrating explicitly how various physical processes influence the interaction between galaxies and Ly$\alpha$ forest transmission features.

\subsubsection{Model: the mean Ly$\alpha$ transmitted flux around galaxies}

The Ly$\alpha$ optical depth around galaxies depends on the density, ionisation, and thermal state of the IGM. Using the fluctuating Gunn-Peterson approximation (e.g. \citealt{Becker2015b} for review), the Ly$\alpha$ optical depth is given by
\begin{equation}
\tau_{\alpha}\simeq 11\Delta_b^2\left(\frac{\Gamma_{\rm HI}}{10^{-12}\rm~s^{-1}}\right)^{-1}\left(\frac{T}{10^4\rm~K}\right)^{-0.72}\left(\frac{1+z}{7}\right)^{9/2},
\end{equation}
where $\Delta_b$ is the baryon overdensity, $\Gamma_{\rm HI}$ is the $\HI$ photoionisation rate, $T$ is the temperature of the IGM. For the $\HI$ proximity effect, the primarily quantity of interest is the typical $\HI$ photoionisation rate around a galaxy, $\langle\Gamma_{\rm HI}(r)\rangle$, which is enhanced relative to the mean value in the IGM, $\bar{\Gamma}_{\rm HI}$. By averaging over many sightlines (ensemble averaging over density fluctuations), the mean Ly$\alpha$ transmitted flux around galaxies is given by
\begin{align}
&\langle \exp(-\tau_\alpha(r))\rangle= \nonumber \\
&~~~~\int d\Delta_b P_V(\Delta_b)\exp\left[-\bar{\tau}_\alpha(\bar{\Gamma}_{\rm HI},T)\Delta_b^2\left(\frac{\langle\Gamma_{\rm HI}(r)\rangle}{\bar{\Gamma}_{\rm HI}}\right)^{-1}\right],\label{eq:model}
\end{align}
where 
$\bar{\tau}_{\alpha}(\bar{\Gamma}_{\rm HI},T)\simeq 11\left(\frac{\bar{\Gamma}_{\rm HI}}{10^{-12}\rm~s^{-1}}\right)^{-1}\left(\frac{T}{10^4\rm~K}\right)^{-0.72}\left(\frac{1+z}{7}\right)^{9/2}$ is the optical depth  at mean and $P_V(\Delta_b)$ is the volume-weighted density probability distribution function \citep{Miralda-Escude2000}, for which we use the \citet{Pawlik2009} fitting formula based on the cosmological hydrodynamical simulations. We assume a uniform temperature of $T=10^{4}\rm~K$ as a fiducial value unless otherwise stated, but examine the impact of the IGM temperature later in the paper.

The model embraces a number of physical factors -- density fluctuations, UV background, and thermal state of the IGM -- important for the mean Ly$\alpha$ transmitted flux around LBGs. We discuss each physical process in the following section.
 
 \subsubsection{Balancing the galaxy abundance with the photoionisation rate required by statistical $\HI$ proximity effect}
 
To derive a constraint on the LyC escape fraction from the statistical $\HI$ proximity effect, we balance the observed galaxy number density with the photoionisation rate required from the Ly$\alpha$ transmitted flux. We formulate this cosmological radiative transfer problem using a statistical argument;  the full treatment is presented in Appendix~\ref{app:model} for an interested reader. Here we focus on the physics essential for understanding the workflow of the methodology.
 
Each star-forming galaxy emits LyC photons at the ionising photon production rate \citep{Robertson2013},
\begin{equation}
\dot{N}_{\rm ion}=f_{\rm esc}\xi_{\rm ion}L_{\rm UV},
\end{equation}
where $f_{\rm esc}$ is the LyC escape fraction, the LyC photon production efficiency $\xi_{\rm ion}$ is the ratio of ionising and non-ionising UV photons, and $L_{\rm UV}$ is the non-ionising UV ($1500$~\AA) luminosity (in units of $\rm erg~s^{-1}~Hz^{-1}$). The total ionising photon production rate density (in units of $\rm photons~s^{-1}~cm^{-3}$) is supplied by all star-forming galaxies above a certain minimum UV luminosity $L^{\rm min}_{\rm UV}$,
\begin{equation}
\dot{\bar{n}}_{\rm ion}(>L^{\rm min}_{\rm UV})=\langle f_{\rm esc}\xi_{\rm ion}\rangle \int^\infty_{L^{\rm min}_{\rm UV}}L_{\rm UV}\Phi(L_{\rm UV})dL_{\rm UV},\label{eq:nion}
\end{equation}
where $\langle f_{\rm esc}\xi_{\rm ion}\rangle$ is the population average of the product of the LyC escape fraction and LyC photon production efficiency and $\Phi(L_{\rm UV})$ is the UV luminosity function. $\langle\cdot\rangle$ means the ensemble-averaged quantity.

The UV luminosity function at $z\sim6$ is now well constrained by both {\it Hubble} Ultra Deep Field and Frontier Field data (we adopt the the UV luminosity function of \citet{Bouwens2015}). Thus the primary unknowns are $\langle f_{\rm esc}\rangle$ and $L^{\rm min}_{\rm UV}$. Although the unknown parameter always comes in the product, $\langle f_{\rm esc}\xi_{\rm ion}\rangle$, $\xi_{\rm ion}$ can be derived from SED fitting \citep{Bouwens2016} or UV metal line ratios \citep{Stark2015,Stark2017,Matthee2017,Harikane2017}.

The independent measure of the ionising photon production rate density comes from the mean transmitted flux in the Ly$\alpha$ forest, which provides a measure of the $\HI$-photoionisation rate of the IGM, $\bar{\Gamma}_{\rm HI}$ \citep[e.g.][]{FG2008,Becker2013}, whence:
\begin{equation}
\bar{\Gamma}_{\rm HI}=\int_{\nu_{\rm HI}}^\infty\sigma_{\rm HI}(\nu)\frac{4\pi \bar{J}_\nu}{h\nu}d\nu\simeq\frac{\alpha_{\rm g}}{\alpha_{\rm g}+3}\sigma_{912}\lambda_{\rm mfp}\dot{\bar{n}}_{\rm ion}(>L^{\rm min}_{\rm UV}),
\end{equation}
where $\sigma_{912}=6.35\times10^{-18}\rm~cm^{2}$ is the $\HI$ photonionisation cross section at the Lyman limit and $\alpha_g$ is EUV (>13.6 eV) spectral slope of galaxies. Both the EUV spectral slope $\alpha_{\rm g}$ and the LyC photon production efficiency $\xi_{\rm ion}$ characterise the hardness of the galaxy spectra; for a given population synthesis model \citep[e.g.][]{Bruzual2003,BPASS17} the best-fit SED fixes both $\alpha_{\rm g}$ and $\xi_{\rm ion}$. We use the mean free path of ionising photons provided by \citet{Worseck2014}, $\lambda_{\rm mfp}\simeq 6.0[(1+z)/7]^{-5.4}\rm~pMpc$. 

In previous work, \citet{Becker2013} \citep[see also][]{Inoue2006,Kuhlen2012} have used the {\it global mean} of the photoionisation rate from Ly$\alpha$ forest at $2<z<5$ and the observed UV luminosity function of galaxies to derive $\langle f_{\rm esc}\rangle$ at a given $L_{\rm UV}^{\rm min}$. Applying this {\it global mean} method, however, becomes difficult at $z>5$ because of the large spatial fluctuations in the intergalactic opacity of the IGM \citep{Becker2015a,Bosman2018}. The local UV background may differ from the global mean, therefore hindering any balance between the mean galaxy number density and the global mean of the UV background.

The statistical $\HI$ proximity effect provides a natural way forward by providing a measure of the local photoionisation rate $\langle\Gamma_{\rm HI}(r)\rangle$ in the same cosmic volume. The average $\HI$-photoionisation rate around a LBG depends on the LyC photons both from a central luminous (detected) LBG and fainter (undetected) galaxies around the central system:
\begin{equation}
\langle\Gamma_{\rm HI}(r)\rangle=\langle\Gamma_{\rm HI}^{\mbox{\tiny LBG}}(r)\rangle+\langle\Gamma_{\rm HI}^{\mbox{\tiny CL}}(r)\rangle.
\end{equation}
The local ionising effect caused by a spectroscopically-detected luminous LBG is 
\begin{equation}
\langle\Gamma_{\rm HI}^{\mbox{\tiny LBG}}(r)\rangle=\frac{\alpha_g \sigma_{912}}{\alpha_g+3}\frac{\langle\dot{N}_{\rm ion}^{\mbox{\tiny LBG}}\rangle}{4\pi r^2}e^{-r/\lambda_{\rm mfp}},
\end{equation}
where $\langle\dot{N}_{\rm ion}^{\mbox{\tiny LBG}}\rangle=\langle f_{\rm esc}\xi_{\rm ion}\rangle\langle L_{\rm UV}\rangle$ is the mean ionising production rate for which the average UV luminosity is given directly from the observed UV magnitudes. Furthermore, the collective LyC photon flux from the fainter undetected galaxies depends on the {\it luminosity-weighted} galaxy correlation function $\langle \xi_g(r)\rangle_L$ (or power spectrum $\langle P_g(k)\rangle_L$) between the luminous LBGs and fainter galaxies above a certain minimum UV luminosity $L^{\rm min}_{\rm UV}$ (see Appendix~\ref{app:model}),
\begin{align}
&\langle\Gamma_{\rm HI}^{\mbox{\tiny CL}}(r)\rangle=\frac{\bar{\Gamma}_{\rm HI}}{\lambda_{\rm mfp}}\int \frac{e^{-|\br-\br'|/\lambda_{\rm mfp}}}{4\pi|\br-\br'|^2}\left[1+\langle\xi_g(|\br'|)\rangle_L\right]d^3r',  \nonumber \\
&~~~~~~~~~~=\bar{\Gamma}_{\rm HI}\left[1+\int_0^\infty \frac{k^2dk}{2\pi^2} R(k\lambda_{\rm mfp})\langle P_g(k)\rangle_L\frac{\sin kr}{kr}\right],\label{eq:GammaCL}
\end{align}
where $R(k\lambda_{\rm mfp})=\arctan(k\lambda_{\rm mfp})/(k\lambda_{\rm mfp})$ is the Fourier transform of the radiative transfer kernel $e^{-r/\lambda_{\rm mfp}}/(4\pi r^2\lambda_{\rm mfp})$. The second equality is succinctly expressed in Fourier space. The luminosity-weighted galaxy power spectrum is
\begin{equation}
\langle P_g(k)\rangle_L=\frac{\int^\infty_{L_{\rm UV}^{\rm min}} L_{\rm UV}\Phi(L_{\rm UV})P_g(k,L_{\rm UV})dL_{\rm UV}}{\int^\infty_{L_{\rm UV}^{\rm min}} L_{\rm UV}\Phi(L_{\rm UV})dL_{\rm UV}},\label{eq:14}
\end{equation}
where  $P_g(k,L_{\rm UV})$ is the Fourier transform of the galaxy correlation function of LBGs with galaxies of luminosity $L_{\rm UV}$. This captures the contribution of galaxies clustered around the LBGs to the ionising background. To estimate the galaxy power spectrum, we use the conditional luminosity function (CLF) approach to populate dark matter halos with galaxies \citep{Yang2003,vandenBosch2013} described fully in Appendix~\ref{app:model}.  The CLF model is constrained by simultaneously fitting the UV luminosity function of $z\sim6$ LBGs from {\it Hubble} Legacy Fields  \citep{Bouwens2015} and the LBG angular correlation function from the {\it HST}+Subaru/Hyper Suprime-Cam samples \citep{Harikane2016}.

Note that the LyC escape fraction enters as $\langle \Gamma_{\rm HI}(r)\rangle\propto \langle f_{\rm esc}\xi_{\rm ion}\rangle$. To see the parameter dependence, it is informative to schematically write:
\begin{equation}
\langle\Gamma_{\rm HI}(r)\rangle\propto\langle f_{\rm esc}\rangle\times \frac{\alpha_{\rm g} {\langle \xi_{\rm ion}\rangle}}{\alpha_{\rm g}+3}\times
\begin{bmatrix} 
	\mbox{\it Galaxy abundance:} \\ \mbox{\it {\small LBG} + galaxy clustering \small{$P_g(k)$}}
\end{bmatrix},
\end{equation}
where we assumed $f_{\rm esc}$ and $\xi_{\rm ion}$ are statistically independent. This highlights how a measure of $\langle\Gamma_{\rm HI}(r)\rangle$ from the statistical $\HI$ proximity effect is balanced with the galaxy abundance estimate from the luminosity function and angular clustering measurements, leading to a constraint on the product of LyC escape fraction and ionising photon production efficiency.

Noting the spectral hardness of ionising sources enters as a combination of the EUV slope and ionising production efficiency, we define an {\it effective spectral hardness parameter} $\langle\xi^{\rm eff}_{\rm ion}\rangle$ and assume a fiducial value,
\begin{equation}
\log \langle\xi^{\rm eff}_{\rm ion}\rangle / [\rm erg^{-1}Hz]=\log\left(\frac{\alpha_{\rm g} {\langle \xi_{\rm ion}\rangle}}{\alpha_{\rm g}+3}\right)=24.8~(fiducial).
\end{equation}
We have adopted a canonical value for the ionising photon production efficiency, $\log\xi_{\rm ion}/[\rm erg^{-1}Hz]=25.2$ \citep{Robertson2013} consistent with LBG observations at intermediate redshift \citep{Bouwens2016,Shivaei2017}. The EUV slope varies from $\alpha_{\rm g}=1$ to $3$ \citep{Kuhlen2012,Becker2013} depending on metallicity and age \citep{BPASS17}. For simplicity, we adopt a fiducial value of $\alpha_{\rm g}=2$. However, adopting $\alpha_{\rm g}=1-3$ only changes the value of $\langle\xi^{\rm eff}_{\rm ion}\rangle$ by 0.2 dex, comparable to the typical uncertainty.

In this radiative transfer model, the nominal free parameters of interest are the product of the LyC escape fraction and LyC photon production efficiency, $\langle f_{\rm esc}\xi_{\rm ion}\rangle$, and the minimum UV luminosity of galaxies that contribute to reionisation, $L_{\rm UV}^{\rm min}$. We vary both parameters when fitting the model to the observed mean Ly$\alpha$ transmitted flux around LBGs, thereby deriving a constraint on the LyC escape fraction. Before presenting the derived constraint on the LyC escape fraction from the statistical $\HI$ proximity effect, we first discuss the impacts of individual physical processes on the mean Ly$\alpha$ transmitted flux around galaxies. 

\subsection{Physical processes governing the mean Ly$\alpha$ transmitted flux around galaxies}\label{sec:physical_processes}

The spatial relationship between galaxies and Ly$\alpha$ forest features carries a wealth of information about the physics of early galaxy formation and reionisation.

\subsubsection{UV background}

Although the UV background includes a contribution from those luminous LBGs detected in our DEIMOS survey, such central LBGs have little impact on the large-scale (>1pMpc) mean Ly$\alpha$ transmitted flux around the LBGs. Their average UV luminosity is $\langle L_{\rm UV}^{\rm LBG}\rangle=1.9\pm0.86\times10^{29}\rm~erg~s^{-1}~Hz^{-1}$ where the error indicates the $1\sigma$ scatter of luminosities. The local ionising effect is then
\begin{equation}
\langle\Gamma_{\rm HI}^{\mbox{\tiny LBG}}(r)\rangle\approx
6.4\times10^{-15} r_{\rm pMpc}^{-2}\left(\frac{\langle f_{\rm esc}\rangle\times\langle\xi^{\rm eff}_{\rm ion}\rangle}{0.1\times10^{24.8}{\rm~erg^{-1}Hz}}\right)\rm~{s^{-1}},
\end{equation}
for $r\ll\lambda_{\rm mfp}$ and $r_{\rm pMpc}=r/(1\rm~pMpc)$ is a distance from the central LBG in proper Mpc. This is more than one order of magnitude lower than the $z\sim6$ mean photoionisation rate measurement from the mean Ly$\alpha$ transmitted flux of the IGM $\bar{\Gamma}_{\rm HI}=1.8^{+1.8}_{-0.9}\times10^{-13}\rm~s^{-1}$ \citep{Wyithe2011}. The same would be true even if the ionising radiation were harder $\log_{10}\xi_{\rm ion}/{\rm [erg^{-1}~Hz]}=25.6$ \citep[e.g.][]{Stark2017} or if we assume a LyC escape fraction of unity. This demonstrates that fainter galaxies, undetected in our survey, are needed to explain the large-scale statistical $\HI$ proximity effect. In Figure \ref{fig:model} the contribution of these fainter galaxies is shown for different values of the mean LyC escape fraction $\langle f_{\rm esc}\rangle$ and the minimum UV luminosity $L^{\rm lim}_{\rm UV}$ (or $M_{\rm UV}^{\rm lim}$) assuming the observed $z\sim6$ UV luminosity function \citep{Bouwens2015} and angular clustering \citep{Harikane2016} brighter than $M_{\rm UV}^{\rm lim}$  (see Appendix \ref{app:model}).  A higher escape fraction increases the average photoionisation rate, enhancing the strength of the statistical $\HI$ proximity effect. Integrating to a fainter $M_{\rm UV}^{\rm lim}$ clearly has a similar effect.

The radial dependence of the Ly$\alpha$ transmitted flux, however, provides additional information on the clustering bias of ionising sources, which, in principle, offers a means to break the degeneracy between $\langle f_{\rm esc}\rangle$ and $M_{\rm UV}^{\rm lim}$. Figure \ref{fig:model} ({\it right}) shows that if only bright galaxies reionise the IGM, they will be clustered more strongly, producing a somewhat steeper slope of the average photoionisation rate and mean Ly$\alpha$ transmitted flux. However, if faint galaxies dominate reionisation (extending below the current {\it Hubble} UV magnitude limit $\approx-15$, e.g. \citealt{Bouwens2017}), their weaker clustering will produce a flatter slope. The luminosity-weighted bias can easily be modelled: on the large scale $\langle P_g(k)\rangle_L\approx b_{\mbox{\tiny LBG}}\langle b_g\rangle_L P_m(k)$ we have
\begin{align}
&\langle\Gamma_{\rm HI}^{\mbox{\tiny CL}}(r)\rangle\approx \nonumber \\
&~~~\bar{\Gamma}_{\rm HI}\left[1+b_{\mbox{\tiny LBG}}\langle b_g\rangle_L\int_0^\infty \frac{k^2dk}{2\pi^2} R(k\lambda_{\rm mfp}) P_m(k)\frac{\sin kr}{kr}\right],
\end{align}
where $\langle b_g\rangle_L$ is the luminosity-weighted bias factor\footnote{Note that the luminosity-weighted bias factor is typically much larger than the normal bias factor \citep{2016MNRAS.457.3541C}, contributing to a large spatial cross-correlation.} of ionising galaxies above $L_{\rm UV}^{\rm min}$:
\begin{equation}
\langle b_g\rangle_L=\frac{\int_{L_{\rm UV}^{\rm min}}^\infty L_{\rm UV}b_g(L_{\rm UV})\Phi(L_{\rm UV})dL_{\rm UV} }{\int_{L_{\rm UV}^{\rm min}}^\infty L_{\rm UV}\Phi(L_{\rm UV})dL_{\rm UV}},
\end{equation}
and $b_{\mbox{\tiny LBG}}$ is the bias factor of LBGs ($M_{\rm UV}<-21$) and $b_g(L_{\rm UV})$ is the bias factor of  galaxies with luminosity $L_{\rm UV}$. The constraint on $\langle b_g\rangle_L$ from the observed mean Ly$\alpha$ transmitted flux around LBGs can thus be translated to a measure of the minimum UV luminosity once combined with the galaxy luminosity function $\Phi(L_{\rm UV})$ and angular correlation function measurements (i.e. $b_g(L_{\rm UV}$)). 

The mean free path $\lambda_{\rm mfp}$ of ionising photons also impacts the radial dependence of the Ly$\alpha$ transmitted flux by setting the maximum distance for influencing the IGM. It is controlled by the number density of $\HI$ absorbers, primarily  Lyman-limit systems. Our assumed value at $z\sim6$ value is based on an extrapolation of the trend within $2.3<z<5.5$ \citep{Worseck2014}. However, hydrodynamical simulations predict $\lambda_{\rm mfp}$ falls markedly at the  end of reionisation \citep{Gnedin2006,Rahmati2017}.  A further uncertainty may arise if Lyman-limit systems are clustered around galaxies; \citet{Rudie2013} find that inclusion of the CGM of galaxies reduces $\lambda_{\rm mfp}$ by $20~\%$. Ultimately, the galaxy-Ly$\alpha$ forest cross-correlation analysis of many QSO sightlines should be interpreted with detailed hydrodynamical simulations. In this analysis, we quantify this modelling uncertainty by lowering $\lambda_{\rm mfp}$ by 20 per cent (i.e. $\lambda_{\rm mfp}=4.8$ pMpc)  for a comparison. 

\subsubsection{Gas density fluctuations}

The inhomogeneous gas distribution in the IGM has the effect of rendering individual associations between galaxies and Ly$\alpha$ transmission spikes {\it stochastic}. The Ly$\alpha$ optical depth at the end of reionisation, e.g. at $z=5.8$, is large:
\begin{equation}
\tau_\alpha\approx48\Delta_b^2\left(\frac{\Gamma_{\rm HI}}{2\times10^{-13}\rm~s^{-1}}\right)^{-1}.
\end{equation}
The level of photoionisation rate required by the statistical $\HI$ proximity effect is $\langle\Gamma_{\rm HI}(r)\rangle\approx3.1-1.6\times10^{-13}\rm~s^{-1}$ at radius $r=1-6\rm~pMpc$ (see Figure \ref{fig:model}), corresponding to the Ly$\alpha$ optical depth value of $\tau_\alpha\approx32-61$. Thus, observable Ly$\alpha$ transmission spikes only occur within {\it IGM underdensities ($\Delta_b<1$) even if the UV background is enhanced}. The required gas underdensity for producing a Ly$\alpha$ transmission spike larger than $F_\alpha^{\rm th}(=e^{-\tau_\alpha^{\rm th}})$ is
\begin{equation}
\Delta_b<\Delta_b^{\rm th}=0.25\left(\frac{\tau_\alpha^{\rm th}}{3}\right)^{1/2}\left(\frac{\Gamma_{\rm HI}}{2\times10^{-13}\rm~s^{-1}}\right)^{1/2},
\end{equation}
where $\tau_\alpha^{\rm th}$ is the corresponding pixel optical depth threshold. For a typical identifiable Ly$\alpha$ transmission spike in the Q1148 spectrum (i.e. $\tau_\alpha^{\rm th}=3$ corresponding to a height $F_\alpha\simeq0.05$), using the density fluctuations from cosmological simulations \citep{Pawlik2009}, the expected occurrence probability of Ly$\alpha$ transmission spike is found as 
\begin{equation}
P(<\Delta_b^{\rm th})=\int_0^{\Delta_b^{\rm th}}P_V(\Delta_b)d\Delta_b\simeq8.7~\%
\end{equation}
at $r=1$ pMpc at an enhanced UV background of $\langle\Gamma_{\rm HI}(r)\rangle\approx3.1\times10^{-13}\rm~s^{-1}$ decreasing to $\simeq1.5~\%$ at large distance (for $\langle f_{\rm esc}\rangle=0.1$ and $M_{\rm UV}^{\rm lim}=-15$). The remaining $\gtrsim90~\%$ of the IGM produces opaque Gunn-Peterson troughs even with an enhanced UV background.  Thus, this provides a natural interpretation for the non-exact alignment (see Figure \ref{fig:map}) between a LBG redshift and the nearest Ly$\alpha$ transmission spike. While the enhanced UV background increases the {\it probability} that the Ly$\alpha$ transmission spikes occur at the IGM around LBGs, but the exact location prefers an underdense IGM.

At smaller radii $\lesssim$ 1 pMpc approaching the CGM regime, the gaseous overdensity increases. This counteracts with the UV background as $\tau_\alpha\propto\Delta_b^2\Gamma_{\rm HI}^{-1}$ introducing more absorption and eventually a negative signal in the cross-correlation\footnote{As the probability distribution function $P_V(\Delta_b)$ adopted here is measured from the entire simulation box \citep{Pawlik2009}, the effect of a gaseous overdensity around galaxies is ignored in the model.}. In the intermediate redshift range $z\simeq2-3$, overdensity around LBGs dominates the small-scale mean Ly$\alpha$ transmitted flux \citep{Adelberger2003,Adelberger2005,Crighton2011,Rudie2012,Rakic2012,Tummuangpak2014,Turner2014,Bielby2017}, consistent with a wide range of cosmological hydrodynamical simulations \citep{Rahmati2015,Meiksin2015,Meiksin2017,Turner2017,Sorini2017}\footnote{At scales less than $\sim$100 pkpc, galactic feedback and hydrodynamic processes complicate the distribution of cold gas.}. However the scale where this downturn occurs is $r\lesssim1.5$ pMpc \citep{Turner2014,Bielby2017}, i.e. several times the commonly-defined  CGM scale ($\simeq300$ pkpc). In Appendix~\ref{app:B}, using the linear theory model we show that the effect of galaxy-gas density correlation is below 10-20 per cent level at $\gtrsim$1 pMpc. Given the range we can measure in the Q1148 field, we therefore expect such small-scale effects to be unimportant.

\begin{figure}
\includegraphics[width=1.0\columnwidth]{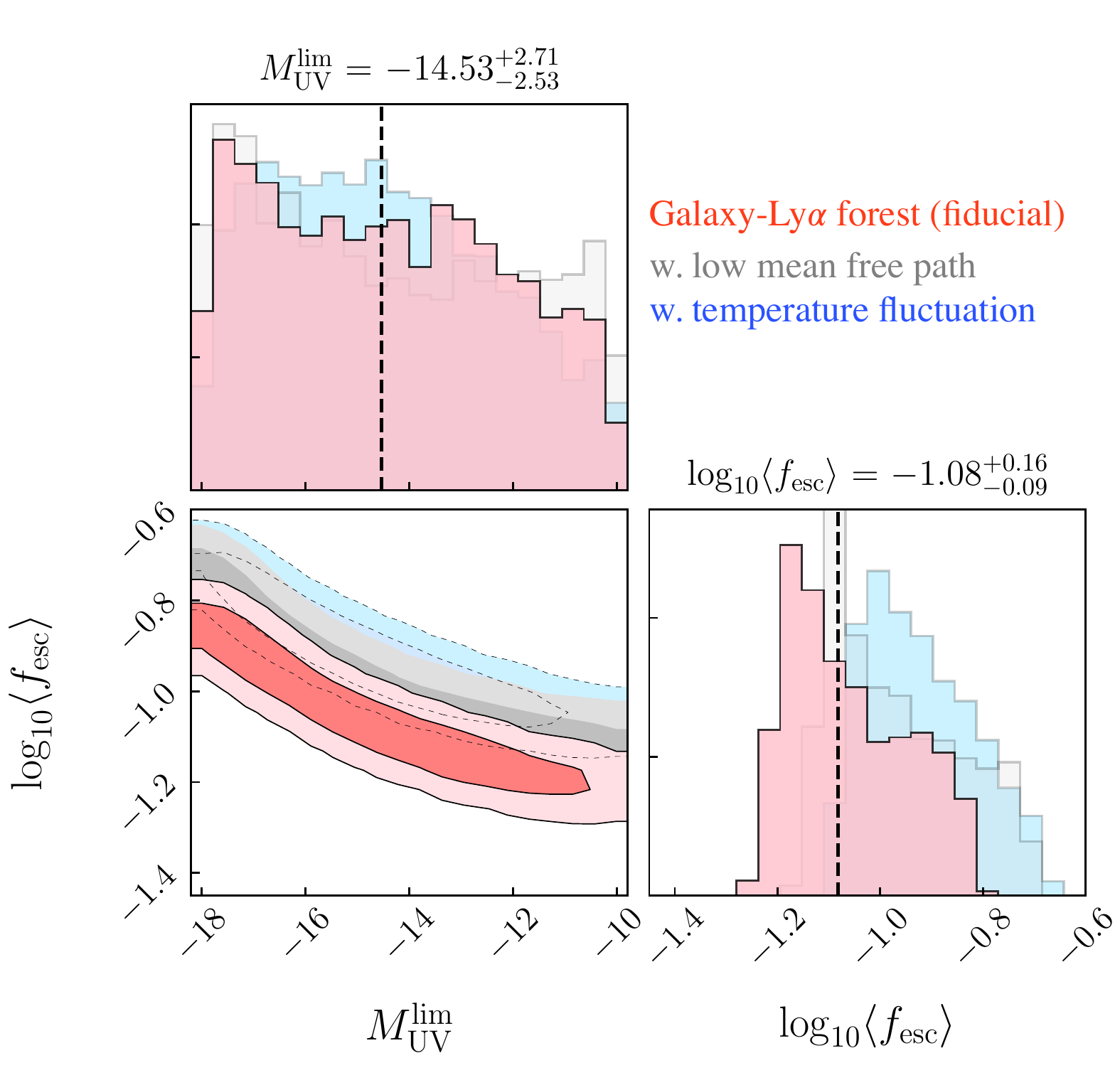}
\vspace{-0.5cm}
\caption{Constraints on the average LyC escape fraction $\langle f_{\rm esc}\rangle$ and the minimum UV luminosity $M_{\rm UV}^{\rm lim}$ with 68 per cent and 95 per cent confidence intervals for the fiducial galaxy-Ly$\alpha$ forest cross-correlation analysis (red: $\lambda_{\rm mfp}=6~\rm pMpc$, $T=10^4\rm~K$) and with a lower value of mean free path (gray: $\lambda_{\rm mfp}=4.8~\rm pMpc$) and with a temperature-density relation (blue: $T_0=10^{4}\rm~K$, $\gamma=1.3$). The quoted constraint is from the fiducial analysis.}\label{fig:MCMC}
\end{figure}

\subsubsection{Thermal state of the IGM}

Thermal fluctuations of the IGM will introduce further modulation of the Ly$\alpha$ optical depth as $\tau_\alpha\propto\Delta_b^2\Gamma_{\rm HI}^{-1}T^{-0.72}$, causing the IGM to be more transparent at higher gas temperature. 
The thermal state of the IGM is primarily controlled by the balance between photoionisation heating and the cooling by adiabatic expansion and Compton scattering off CMB photons; it produces a tight asymptotic power-law relation \citep{Hui1997,McQuinn2016}
\begin{equation}
T=T_0\Delta^{\gamma-1}.
\end{equation}
For $T_0=10^4\rm~K$ and assuming $\gamma=1.3$, the Ly$\alpha$ transmitted flux is lower than for the fiducial $\gamma=1$. This is because the temperature of the underdense IGM which gives rise to Ly$\alpha$ transmission spikes is lower (e.g. $\log_{10}T/{\rm~K}=3.82$ at $\Delta_b=0.25$). Cosmological radiative transfer simulations find a large scatter around $\gamma=1$ in the temperature-density relation just after the IGM is reionised \citep{Tittley2007,Trac2008,Kakiichi2017,Keating2017}, which is not captured by the single power-law relation. Thus, we adopt an uniform temperature for simplicity for a fiducial analysis, but also repeat the analysis with $T=T_0\Delta^{\gamma-1}$ assuming $T_0=10^4\rm~K$ and $\gamma=1.3$. The increased opacity arising from temperature fluctuations requires more ionising photons to match the statistical $\HI$ proximity effect and hence a higher LyC escape fraction.

Large-scale thermal fluctuations may also be caused by environmental effects in the reionisation process. In `inside-out' reionisation, highly biased regions around luminous galaxies are thought to have ionised earlier, allowing more time for the gas to cool by adiabatic expansion and CMB Compton cooling. This causes the low-density IGM near luminous galaxies to be preferentially cooler \citep{DAloisio2015}, reducing the mean Ly$\alpha$ transmitted flux around LBGs at inner radii \citep{Davies2017}. The extent of this effect is debated \citep[e.g.][]{Keating2017}. For the low-density IGM close to luminous galaxies, the temperature asymptotically relaxes to the value set by the balance between the adiabatic expansion and instantaneous photoionisation rate. \ignore{resulting in $T_{\rm asympt.}\approx9400{\rm~K}(\langle E_{\rm HI}\rangle/{5\rm~eV})^{3/5}[(1+z)/4]^{9/10}$ \citep{McQuinn2016} {\bf (KK: check number)}.} On the other hand, the IGM away from the galaxies that has been engulfed by a $\HII$ I-front raises the temperature to about $\sim10^4\rm~K$. The large-scale thermal fluctuations vary from $\sim5000\rm~K$ to $T\approx1.0-1.5\times10^4\rm~K$ which contributes to the negative correlation of the mean Ly$\alpha$ transmitted flux around LBGs. As the temperature has a weaker dependence on the optical depth $\tau_\alpha\propto \Delta_b^2\Gamma_{\rm HI}^{-1}T^{-0.72}$, this can easily be compensated by only moderate enhancement of the UV background. Although both UV background and thermal fluctuations co-exist, because of the steeper dependence on the photoionisation rate it is likely that the UV background variation dominates creating a positive correlation, with secondary modulation by thermal fluctuations weakening it (\citealt{Davies2017}, private communication).

\section{Constraining the mean escape fraction}\label{sec:fesc}

We now utilise the foregoing to analyse the balance between inferred galaxy abundance in the Q1148 field with the observed mean Ly$\alpha$ transmitted flux in terms of a statistically-averaged LyC escape fraction $\langle f_{\rm esc}\rangle$. To
accomplish this we fit the model to the observed mean Ly$\alpha$ transmitted flux data using a Markov Chain Monte Carlo method \citep{Foreman-Mackey2013} varying $\langle f_{\rm esc}\rangle$ and $\Muv^{\rm lim}$. We assume a Gaussian likelihood and place a flat prior in the range of $-2<\log_{10}\langle f_{\rm esc}\rangle<0$ and $-18<M_{\rm UV}^{\rm lim}<-10$. We have tested the result against an enlarged prior range ($-20<M_{\rm UV}^{\rm lim}<-8$) and find a consistent result. For the covariance matrix we only use diagonal elements from the Jackknife error estimate. 

\begin{figure}
\centering
\includegraphics[width=1.0\columnwidth]{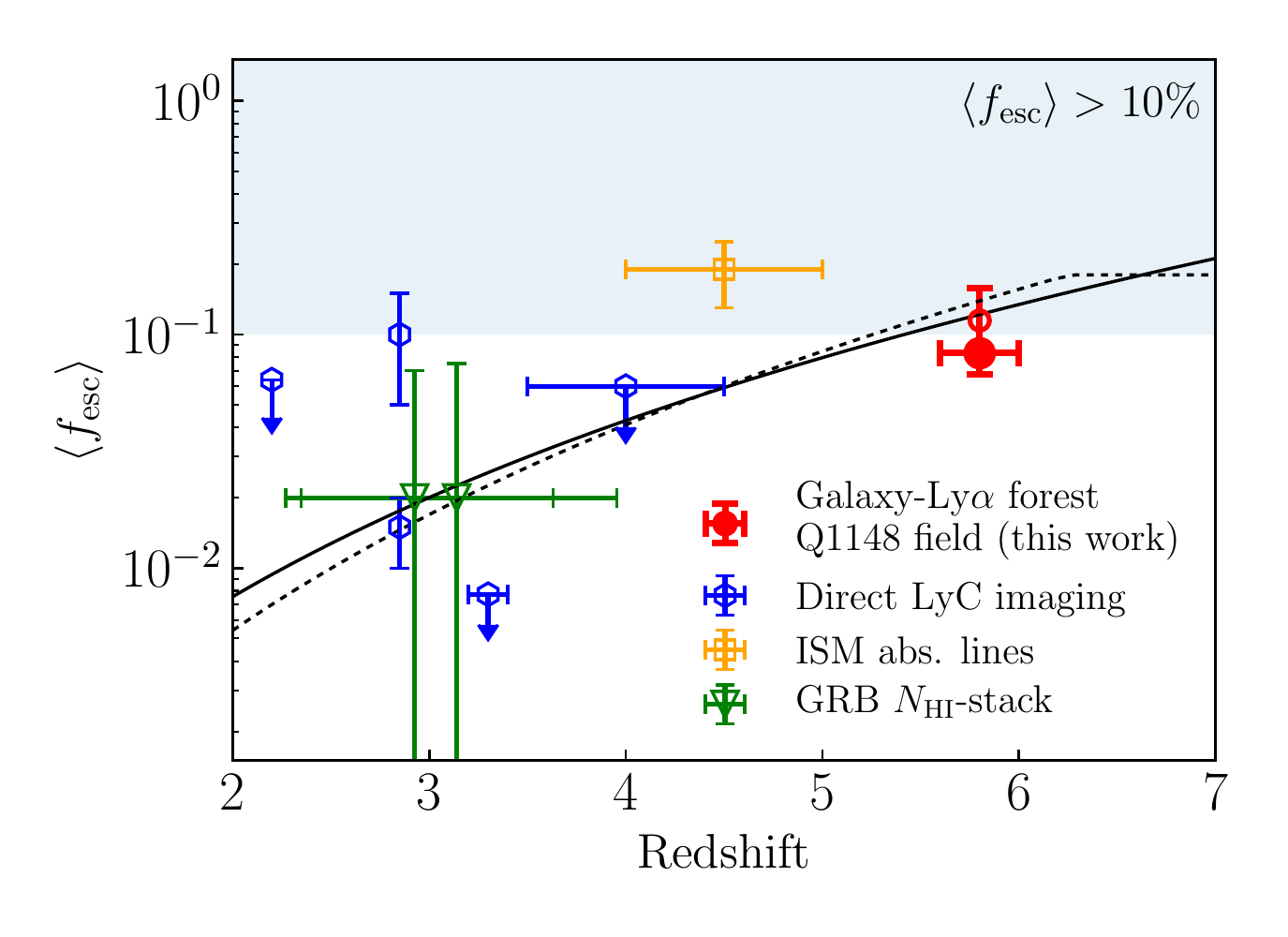}
\vspace{-0.5cm}
\caption{Redshift evolution of the population-averaged LyC escape fraction of galaxies. The $z\simeq6$ constraint from the galaxy-Ly$\alpha$ forest cross-correlation in Q1148 field is indicated by the filled red circle. A compilation of previous $2<z<4$ constraints  is indicated by open symbols. These include direct LyC imaging \citep{Vanzella2010,Mostardi2013,Grazian2016,Matthee2017} and GRB $\NHI$-stacking \citep{Chen2007,Fynbo2009}, and ISM absorption line studies \citep{Leethochawalit2016}. The model mean LyC escape fractions adopted by \citet{HM2012} (solid) and \citet{Puchwein2018} (dotted) are overlaid. The shaded region indicate $\langle f_{\rm esc}\rangle>10\%$ required for galaxies to drive reionisation.}\label{fig:fesc}
\end{figure}

In Figure \ref{fig:MCMC} we show the derived constraint on the $\langle f_{\rm esc}\rangle-M_{\rm UV}^{\rm lim}$ plane. The inferred mean LyC escape fraction at $z\simeq6$ is found to be
\begin{equation}
\langle f_{\rm esc}\rangle =0.083^{+0.037}_{-0.016}\left(\frac{\langle\xi^{\rm eff}_{\rm ion}\rangle}{10^{24.8}\rm~erg^{-1}Hz}\right)^{-1},
\end{equation}
for $\Muv^{\rm lim}=-14.53^{+2.71}_{-2.53}$ for the fiducial analysis.\footnote{Note that for fiducial analysis we have ignored the three radial bins at 3.5-4.5 pMpc as they are likely affected by systematics. Their inclusion would give a 12 \% larger $\langle f_{\rm esc}\rangle$ with two possible best-fit values of $M_{\rm UV}^{\rm lim}$ due to the poor constraint on the shape. } This constraint is dependent upon the assumed mean free path and IGM temperature. Nonetheless, as discussed in the previous section, a lower mean free path and thermal fluctuations would mean a larger (>10 \%) mean LyC escape fraction to compensate the increased opacity. These uncertainties on radiative transfer can be included in the MCMC analysis once a larger dataset becomes available.

Although our sample is modest, our result suggests that $\langle f_{\rm esc}\rangle=0.06-0.16$ for star-forming galaxies above $\Muv^{\rm lim}=-14.53^{+3.16}_{-2.47}$ including modelling systematic error. In Figure \ref{fig:fesc} we compare our $\langle f_{\rm esc}\rangle$ constraint with earlier estimates from LyC imaging at $z\sim2-4$ \citep{Vanzella2010,Mostardi2013,Grazian2016,Matthee2017}. Low escape fractions at $z\sim3$, $\langle f_{\rm esc}\rangle=0.02\pm0.02$ ($<0.075$ at $95\%$ confidence upper limit), are also indicated from $\HI$ covering fractions derived from the spectra of long-duration gamma-ray bursts (GRB) \citep{Chen2007,Fynbo2009}. Our new estimate suggests a rising mean escape fraction with increasing redshifts consistent with the trend adopted by the the recently revised synthesis model of the cosmic UV background \citep{Puchwein2018} and the minimal reionisation model of \citet{HM2012}. This means that faint galaxies deposit sufficient ionising radiation into the IGM for driving the reionisation process (see also \citealt{Faisst2016}). Since the inclusion of temperature fluctuations would require  more ionising photons to match the observed positive correlation of the mean Ly$\alpha$ transmitted flux around LBGs, our fiducial analysis provides a fairly conservative lower limit to the mean LyC escape fraction. 

\section{The Impact of Luminous Systems}\label{sec:ind}

Finally, we turn our attention to two individual cases of a LBG and AGN for which we can identify associated transmission spikes in the Q1148 spectrum. We investigate both as examples of spatial fluctuations in the IGM environment induced by luminous sources. We discuss how they might contribute to spatial fluctuations of the ionisation and thermal states of the IGM and the possible role of rare, luminous sources on the reionisation process.

\subsection{z=6.177 LBG J1148+5250 and Ly$\beta$ transmission spike}

LBG J1148+5250 is a newly-discovered Ly$\alpha$ emitting galaxy in our DEIMOS sample. It is a luminous ($\Muv=-21.8$) galaxy with a secure asymmetric Ly$\alpha$ line at $z_{\rm Ly\alpha}=6.177$. Interestingly, the LBG redshift coincides with that of a Ly$\beta$ transmission spike at $z=6.185$. This is the first case of a possible individual transverse proximity effect around a $z>6$ LBG (Table~\ref{table:LBG}). The Ly$\beta$ transmission spike is separated by $d_{\rm spike}=1.9\rm~pMpc$ ($9.4~h^{-1}\rm cMpc$) from the LBG.

The detection of a Ly$\beta$ transmission spike and the high optical depth in the Ly$\alpha$ forest region (see Figure \ref{fig:LBG-Lyb}) places a bound on the Ly$\alpha$ transmission of the IGM. The peak transmitted flux is $e^{-\tau_{\alpha+\beta}}=0.0686\pm0.0066$ ($\tau_{\alpha+\beta}=2.68$). Because the high redshift ($z>6$) Ly$\beta$ forest overlaps with its lower redshift ($z<5.26$) Ly$\alpha$ equivalent, this translates into an upper limit on the $z=6.185$ Ly$\beta$ optical depth $\tau_\beta<\tau_{\alpha+\beta}$ and, using the ratio between the Ly$\beta$ and Ly$\alpha$ optical depths $\tau_\beta/\tau_\alpha=f_{13}\lambda_\beta/(f_{12}\lambda_\alpha)=0.16$ predicted by atomic physics, a range of
\begin{equation}
4.2~(3\sigma)<\tau_\alpha<16.7\pm0.6,
\end{equation}
consistent with the absence of a clear Ly$\alpha$ transmission spike above the 3$\sigma$ noise in the QSO spectrum. 

\begin{figure}
\hspace{-0.5cm}
\includegraphics[width=1.1\columnwidth]{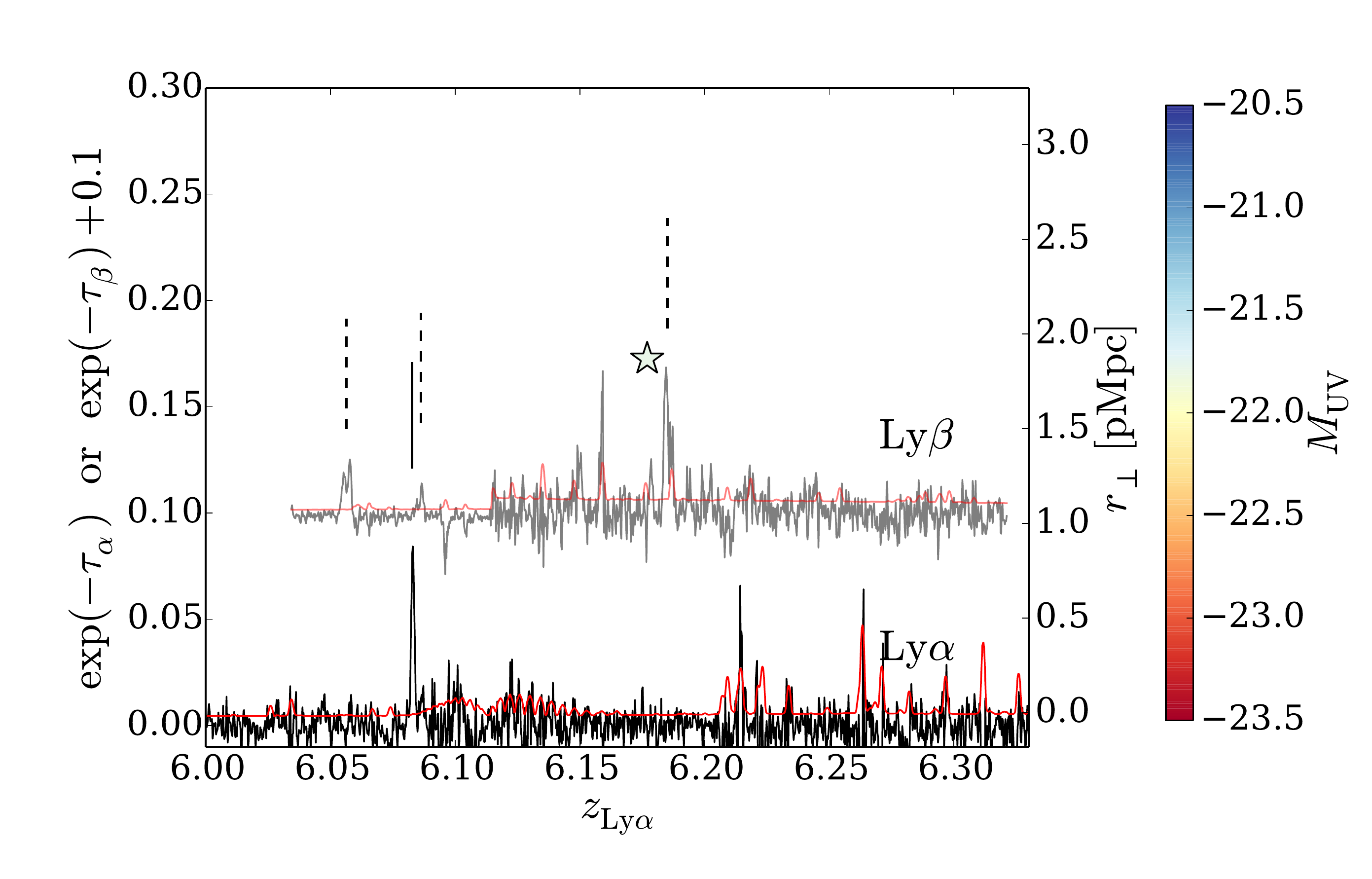}
\vspace{-0.5cm}
\caption{A zoom in of Figure~\ref{fig:map} around the luminous LBG J1148+5250 at $z_{\rm Ly\alpha}=6.177$ adopting the same colour bar for the galaxy luminosity. Solid and dashed vertical lines indicate the location of wavelet-identified Ly$\alpha$ and Ly$\beta$ transmission spikes. The Ly$\beta$ forest region is offset by 0.1 in y-axis.}\label{fig:LBG-Lyb}
\end{figure}

\begin{figure}
\centering
\includegraphics[width=\columnwidth]{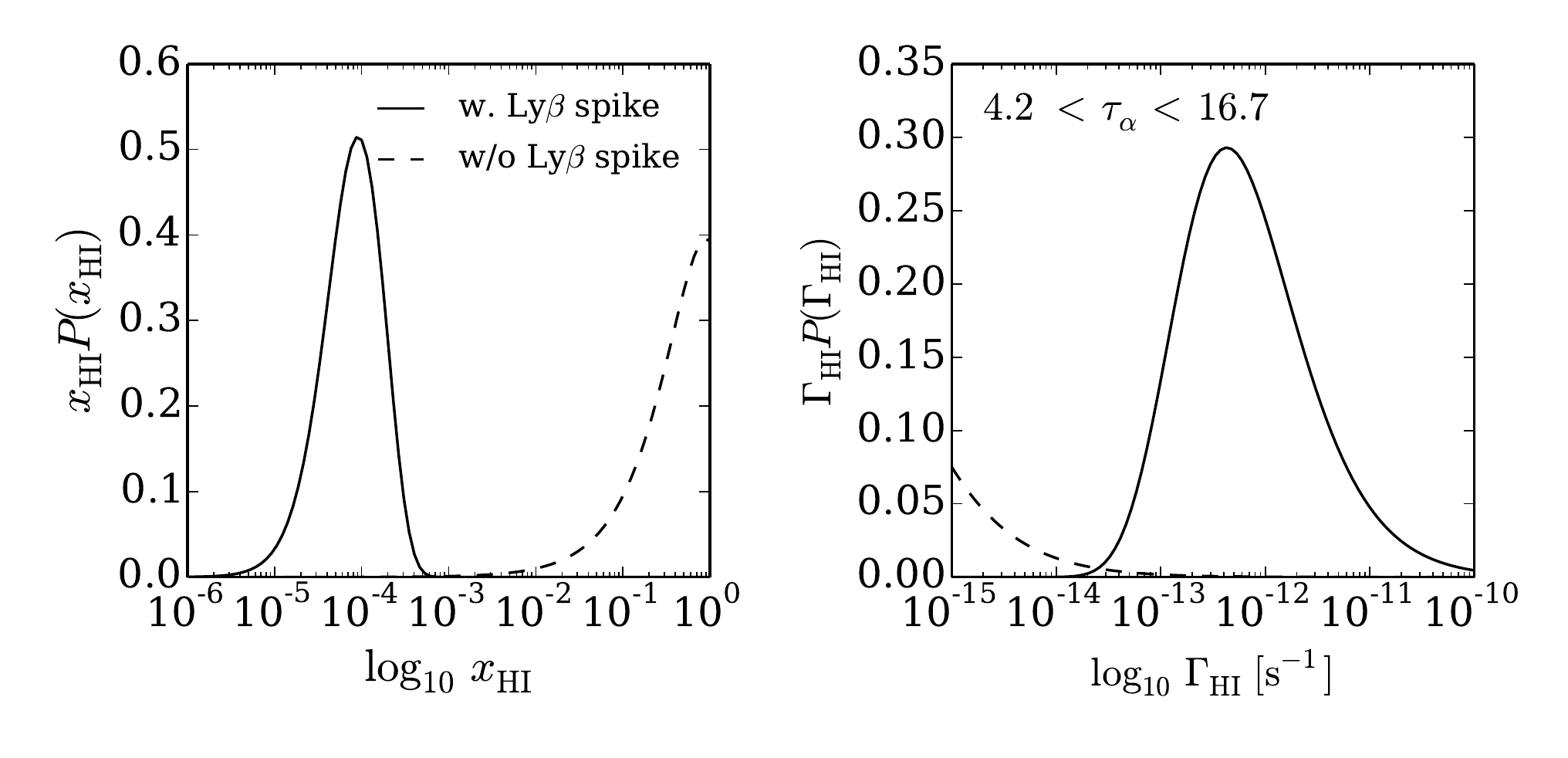}
\vspace{-0.5cm}
\caption{The probability distribution of ({\it left panel}) the neutral hydrogen fraction and ({\it right panel}) the $\HI$ photoionisation rate at the location of the Ly$\beta$ transmission spike $z=6.185$ (solid). For comparison, the dashed line shows a hypothetical case without a Ly$\beta$ transmission spike (assuming the Ly$\alpha$ optical depth can reach the Gunn-Peterson optical depth of a fully neutral medium, $\tau_\alpha=1.8\times10^5$).}\label{fig:Lyb_PDF}
\end{figure}

Compared to the Gunn-Peterson optical depth at $z=6.185$, $\tau_{\rm GP}\simeq1.8\times10^5\xHI\Delta_b$, this upper limit on $\tau_\alpha$ is quite low, suggesting that the IGM is highly ionised to $\xHI\lesssim10^{-4}$. As discussed in Section \ref{sec:physical_processes}, the association of individual galaxies and transmission spikes is probabilistic owing to the gas density fluctuations. Thus, we should assess the probability distribution of the neutral hydrogen fraction $\xHI$ at the location of the Ly$\beta$ transmission spike given an observed Ly$\alpha$ optical depth. Using the simulated probability distribution function of gas density fluctuations and $\tau_\alpha=\bar{\tau}_{\mbox{\tiny GP}}\xHI\Delta_b$ where $\bar{\tau}_{\mbox{\tiny GP}}=1.8\times10^5$ is the Gunn-Peterson optical depth of a fully neutral medium at mean density, we find that 
\begin{equation}
P(\xHI|\tau_\alpha)=\int \delta_D\left(\xHI-\frac{\tau_\alpha}{\bar{\tau}_{\mbox{\tiny GP}}}\Delta_b^{-1}\right)P_V(\Delta_b)d\Delta_b.\label{eq:PDF}
\end{equation} 
Figure \ref{fig:Lyb_PDF} (left) shows the resulting probability distribution of the neutral fraction $\xHI$ after marginalising over the observed bound of the Ly$\alpha$ optical depth. The presence of a Ly$\beta$ transmission spike indeed indicates that the $z=6.185$ IGM is highly ionised to the expected value of $\xHI\simeq10^{-4}$. Note that this analysis does not assume the medium is photoionized a priori. Thus, {\it a UV luminous galaxy at the reionisation epoch ($z>6$) is clearly located in a highly ionized environment}. 

\begin{table}
\centering
\caption{
Summary of the IGM environment of the $z=6.177$ luminous Ly$\alpha$ emitting LBG. The associated Ly$\beta$ transmission spike is the evidence of highly ionised intergalactic gas around the LBG, which is maintained likely by the faint galaxy overdensity (indicated by the excess $\OI$ absorbers).
}
\label{table:LBG}
\begin{tabular}{ll}
\hline\hline
LBG's Ly$\alpha$ redshift & $z=6.177$  \\ 
Ly$\beta$ transmission spike & $z=6.185$  \\ 
Lower limit to the $\HII$ bubble size$^\ast$ & $>1.9\rm~pMpc$ ($9.4h^{-1}\rm cMpc$) \\
Photoionisation rate of the LBG$^\dagger$ & $\Gamma^{\mbox{\tiny{LBG}}}_{\mbox{\tiny HI}}\simeq2.1\times10^{-15}\rm~s^{-1}$ \\
Photoionisation rate at the Ly$\beta$ spike$^\ddagger$ &  $\Gamma^{\mbox{\tiny{spike}}}_{\mbox{\tiny HI}}\simeq5.7\times10^{-13}\rm~s^{-1}$ \\
\begin{tabular}{@{}l@{}} $\OI$ absorbers' redshift \\ (distance to the Ly$\beta$ spike) \end{tabular} &  
\begin{tabular}{@{}l@{}} $z=6.1293, 6.1968, 6.2555$ \\ ~~~~~~~~($3.2, 0.7, 4.0\rm~pMpc$) \end{tabular} \\
\hline
\multicolumn{2}{l}{$^\ast$ From the distance between the LBG and Ly$\beta$ spike.} \\
\multicolumn{2}{l}{$^\dagger$ At the Ly$\beta$ spike (i.e. $1.9\rm~pMpc$ distance from the LBG) and for} \\
\multicolumn{2}{l}{~~~$\rm SFR=28~M_\odot~yr^{-1}$, $f_{\rm esc}=0.1$, and $\xi_{\rm ion}=10^{25.2}\rm erg^{-1}Hz$.}  \\
\multicolumn{2}{l}{$^\ddagger$ The expected median value of the photoionisation rate at the location} \\
\multicolumn{2}{l}{~~~of the Ly$\beta$ spike (see Figure 12).} \\
\end{tabular}
\end{table}

The distance to the Ly$\beta$ transmission spike from LBG J1148+5250 provides a lower limit to the size of the cosmological $\HII$ region,
\begin{equation}
R_{\rm HII}>d_{\rm spike}=1.9{\rm~pMpc}~(9.4~h^{-1}{\rm cMpc})~~~\mbox{at}~~z=6.18.
\end{equation}

Can this luminous galaxy alone produce such a large ionised bubble? The UV luminosity $L_{\rm UV}=2.25\times10^{29}\rm~erg~s^{-1}~Hz^{-1}$ corresponds to a star formation rate ${\rm SFR}=28.1\rm~M_\odot~yr^{-1}$ assuming a Salpeter IMF and solar metallicity \citep{Madau1998} before any correction for dust extinction. We can estimate the size of the $\HII$ region 
\begin{align}
&R_{\rm HII}=\left[\frac{3}{4\pi}\frac{\dot{N}_{\rm ion}^{\HII} t_{\rm age}}{\bar{n}_{\rm H}(z)}\right]^{1/3}, \nonumber \\
&\approx1.0\left[\left(\frac{f_{\rm esc}}{0.1}\right)\left(\frac{\xi_{\rm ion}}{10^{25.2}{\rm~erg^{-1}Hz}}\right)\left(\frac{t_{\rm age}}{300\rm~Myr}\right)\right]^{1/3}\rm pMpc.
\end{align}
assuming a constant star formation history over the median age of UV luminous galaxies ($L>L^*$) at $z\simeq6$ of 
$\simeq200-300\rm~Myr$ \citep{Curtis-Lake2013}. Even for a hard $\log_{10}\xi_{\rm ion}=25.6$, the radius becomes $R_{\rm HII}\approx1.4\rm~pMpc$ below the observed lower limit. It therefore seems necessary to invoke a contribution from fainter galaxies clustered around the luminous LBG.

The probability distribution of the photoionisation rate inside the $\HII$ region can be estimated as in Equation (\ref{eq:PDF}) by integrating the Dirac delta function at $\Gamma_{\rm HI}\propto\tau_\alpha^{-1}\Delta_b^2$ with $P_V(\Delta_b)$. Figure \ref{fig:Lyb_PDF} shows that the expected photoionisation rate at the Ly$\beta$ transmission spike may be as high as $\Gamma_{\rm HI}\simeq10^{-12}-10^{-13}\rm~s^{-1}$, close to the value indicated by the statistical analysis in Section \ref{sec:physical_processes}. Such a high photoionisation rate cannot be maintained by the luminous LBG alone, which contributes up to $\Gamma_{\rm HI}^{\mbox{\tiny LBG}}(r)\approx
7.6-19.0\times10^{-15} (r/1{\rm~pMpc})^{-2}\rm~s^{-1}$ for $f_{\rm esc}=0.1$ and $\log_{10}\xi_{\rm ion}=25.2-25.6$. 

\citet{Becker2006} report the discovery of four $\OI$ absorbers at $z=6.0097, 6.1293, 6.1968, 6.2555$, which indicates the location of low luminosity galaxies \citep{Finlator2013} below the LBT detection limit ($M_{\rm UV}\simeq-21$). The closest $z=6.1968$ $\OI$ absorber is separated by $\simeq$0.7 pMpc from the Ly$\beta$ transmission spike. Such a surprising excess of $\OI$ absorbers near the $z\simeq6.18$ luminous LBG - Ly$\beta$ transmission spike association supports the presence of clustered faint galaxies around the LBG, and their collective ionising contribution.

In summary, the discovery of a Ly$\beta$ transmission spike near the z$\simeq$6.18 LBG further supports the conclusion of our statistical analysis. Accelerated reionisation is likely driven by the collective ionising contribution from fainter galaxies clustered around luminous LBGs, possibly enhanced with a harder ionising spectrum.

\subsection{A Faint AGN and broad Ly$\alpha$ transmission spikes at z$\sim$5.7}

\begin{figure}
\hspace{-0.5cm}
\includegraphics[width=1.1\columnwidth]{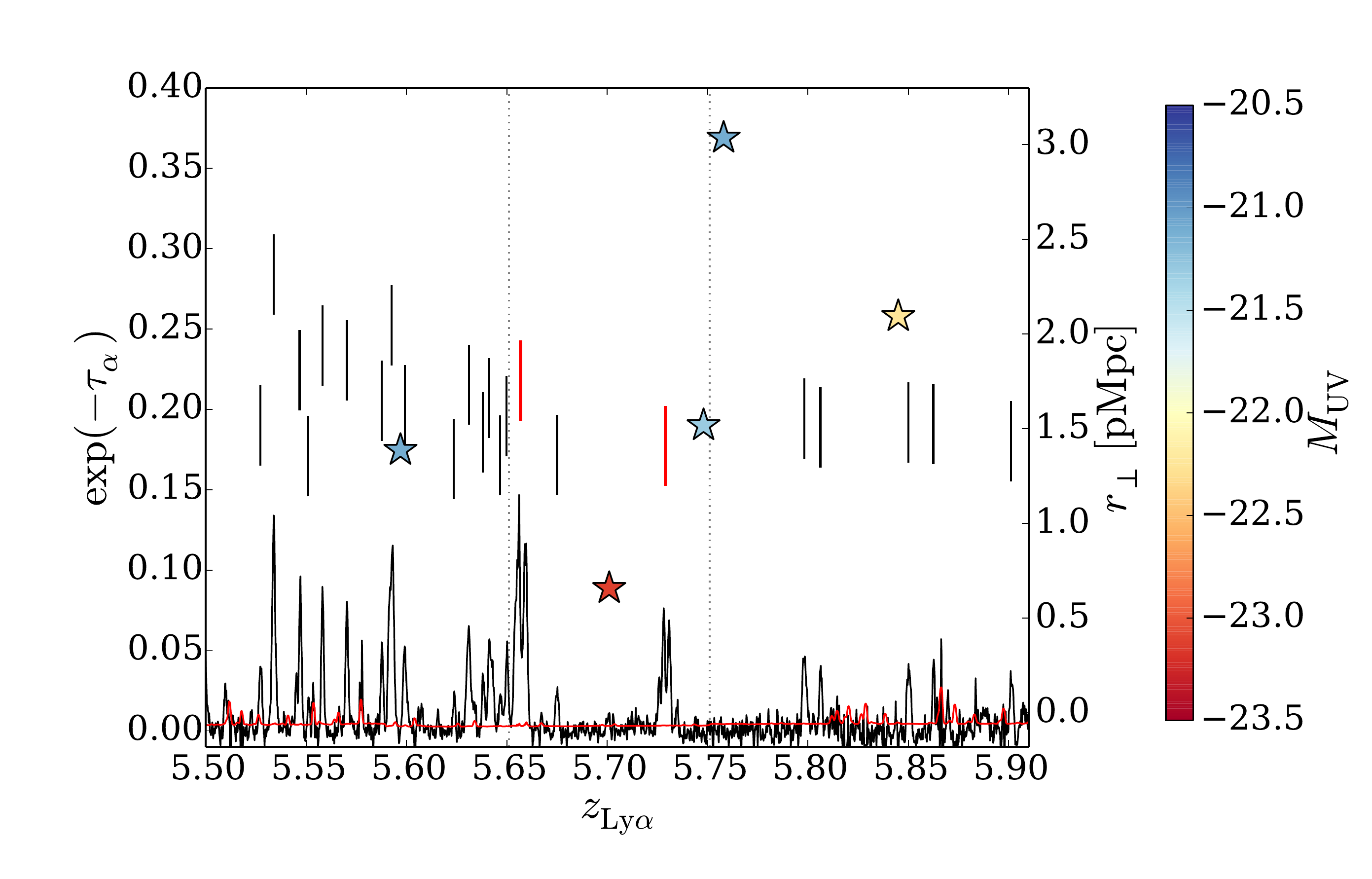}
\vspace{-0.5cm}
\caption{A zoom-in of Figure~\ref{fig:map} around the faint AGN RD J1148+5253 at $z_{\rm Ly\alpha}=5.701$. Solid vertical lines indicates the wavelet-identified Ly$\alpha$ transmission spikes. The red lines mark the two broad Ly$\alpha$ transmission spikes. The $\Delta z=0.1$ ($\approx6.8\rm~pMpc$) region around the AGN is shown by the dotted vertical lines.}\label{fig:AGN-Lya}
\end{figure}

\begin{figure}
\centering
\vspace{-0.5cm}
\includegraphics[width=0.9\columnwidth]{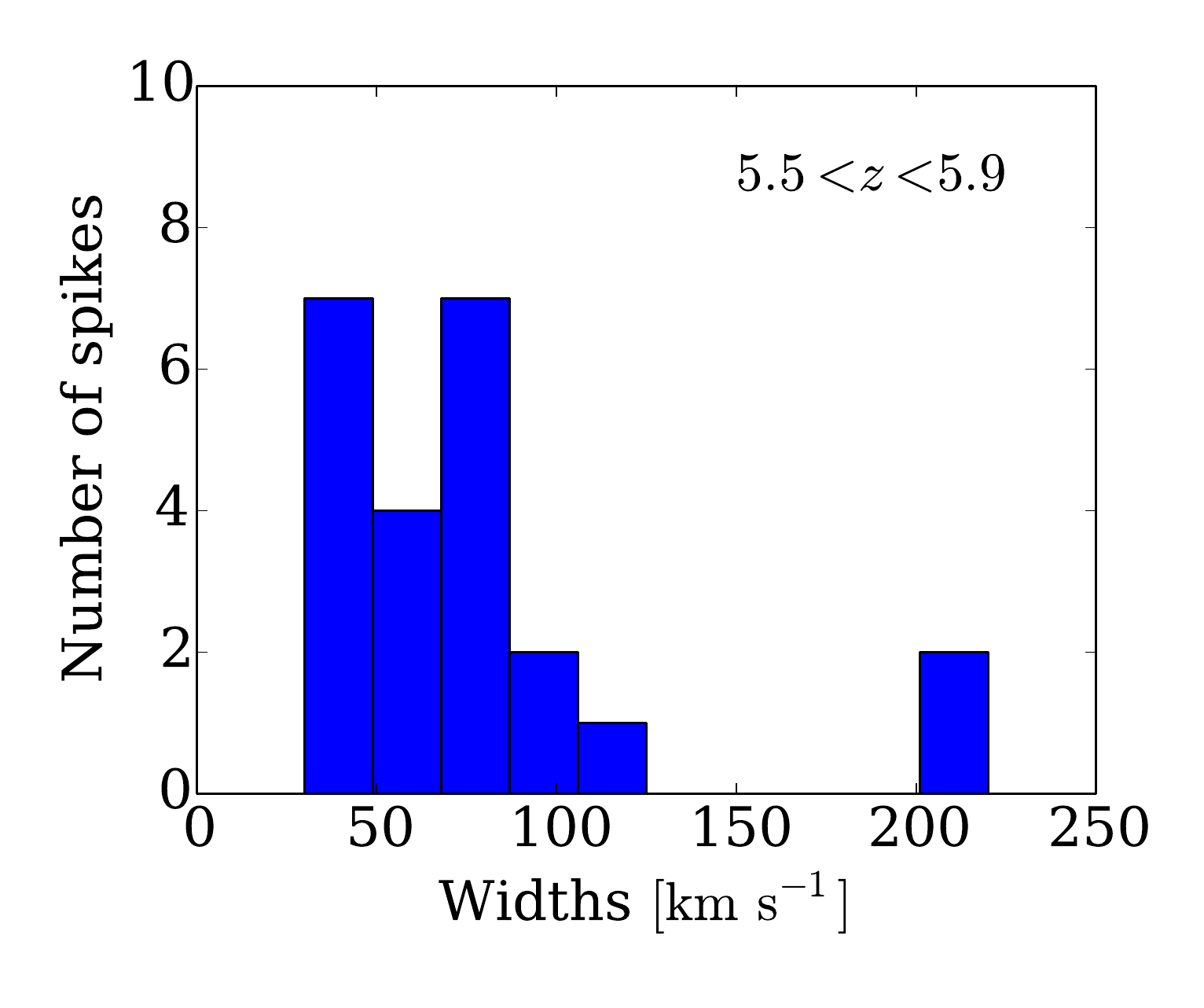}
\vspace{-0.5cm}
\caption{Histogram of the widths of wavelet-identified Ly$\alpha$ transmission spikes in the redshift range of $5.5<z<5.9$. The two values with $\simeq220\rm km~s^{-1}$ width are those indicated by the red lines in Figure \ref{fig:AGN-Lya}.}\label{fig:agn_width}
\end{figure}

RD J1148+5253 is a low luminosity ($\Muv=-23.1$) AGN with a redshift of $z_{\rm Ly\alpha}=5.701$ originally discovered by \citet{Mahabal2005}. We confirm the faint AGN with our deep 4.2 hrs DEIMOS spectroscopy via detection of a broad Ly$\alpha$, $\NV$ $\lambda1240$, Ly$\beta$ emission lines, and the associated continuum (Figure \ref{fig:agn}). Assuming the Eddington luminosity for RD J1148+5253, the super massive black hole (SMBH) mass is estimated to be $M_{\rm BH}\approx5\times10^7\rm~M_\odot$ after applying the bolometric correction of 4.4 \citep{Willott2010} to the observed UV luminosity. Non-thermal emission is evident from the high ionisation metal line $\NV$ as well as from the tentative $2.6-2.9\sigma$ detection of X-ray emission from 78 ks {\it Chandra} observation \citep{Gallerani2017}. 

In the Mpc-scale environment around the AGN (see Figure \ref{fig:AGN-Lya} and Table~\ref{table:AGN}), the spectrum of QSO J1148+5251 exhibits two prominent broad Ly$\alpha$ transmission spikes at $z_{\rm Ly\alpha}=5.729$ and $5.657$ located at 2.02 pMpc and 3.07 pMpc away from the AGN, respectively. Figure \ref{fig:agn_width} shows that the widths of both transmission spikes are broader ($\simeq220~\rm km~s^{-1}$) than others ($\lesssim110~\rm km~s^{-1}$) in the redshift range $5.5<z<5.9$. Could the faint AGN impact the physical origin of the broad Ly$\alpha$ transmission spikes?

The hard ionising spectra of an AGN will enhance the local UV background. Assuming a broken power-law spectrum $L_\nu\propto\nu^{-0.5}$ for $1050$~\AA$~<\lambda<1450$~\AA~and $L_\nu\propto\nu^{-1.5}$  for $\lambda<1050$~\AA~\citep[e.g.][]{Telfer2002}, the $\HI$ ionising photon production rate of RD J1148+5253 is $\dot{N}_{\rm ion}^{\rm HII}=\int^\infty_{\nu_{\rm HI}}L_\nu/(h\nu)d\nu\approx5.3\times10^{55}\rm~s^{-1}$, providing the photoionisation rate of
\begin{equation}
\Gamma_{\rm HI}^{\rm AGN}=\frac{\alpha_Q}{3+\alpha_Q}\frac{\sigma_{912}\dot{N}_{\rm ion}^{\rm HII}}{4\pi r^2}\approx1.0\times10^{-13}\left(\frac{r}{3~\rm pMpc}\right)^{-2}\rm~s^{-1},
\end{equation}
for $\alpha_Q=1.5$. At the location of the broad Ly$\alpha$ transmission spikes ($r=2-3$ pMpc), the faint AGN alone gives an optical depth $\tau_\alpha\approx48-110\Delta_b^2(T/10^4{\rm~K})^{-0.72}$. While the ionising contribution of the faint AGN is somewhat larger than a luminous LBG, once again in order to match the observed spikes ($\tau_\alpha\simeq2-3$), either a gaseous underdensity ($\Delta_b\lesssim0.25$) or associated fainter ionising sources are required.

\begin{table}
\centering
\caption{
Summary of the IGM environment of the $z=5.701$ faint AGN. The associated two broad Ly$\alpha$ transmission spikes may be due to the impact of the AGN on the ionisation and thermal state of the IGM. 
}
\label{table:AGN}
\begin{tabular}{ll}
\hline\hline
AGN's Ly$\alpha$ redshift & $z=5.701$  \\ 
\begin{tabular}{@{}l@{}} Broad Ly$\alpha$ transmission spikes \\ (distance to the AGN) \end{tabular} & 
\begin{tabular}{@{}l@{}} $z=5.657, 5.729$ \\ ~~~~~~~~($3.07, 2.02\rm~pMpc$) \end{tabular}  \\ 
Photoionisation rate of the AGN$^\dagger$ & $\Gamma^{\mbox{\tiny{LBG}}}_{\mbox{\tiny HI}}\simeq1.0, 2.2\times10^{-13}\rm~s^{-1}$ \\
Photoionisation rate at the Ly$\alpha$ spikes$^\ddagger$ &  $\Gamma^{\mbox{\tiny{spike}}}_{\mbox{\tiny HI}}\simeq8.2, 7.0\times10^{-13}\rm~s^{-1}$ \\
An estimated $\HeIII$ bubble size & $3.2{\rm~pMpc}(t_{\mbox{\tiny Q}}/10^8\rm yr^{-1})$ \\
\hline
\multicolumn{2}{l}{$^\dagger$ At the broad Ly$\beta$ spikes,  $3.07, 2.02\rm~pMpc$ respectively.} \\
\multicolumn{2}{l}{~~~We assumed the 100 per cent escape fraction of the AGN.} \\
\multicolumn{2}{l}{$^\ddagger$ The expected median value of the photoionisation rate at the location} \\
\multicolumn{2}{l}{~~~of the broad Ly$\alpha$ spikes at $3.07, 2.02\rm~pMpc$ respectively.} \\
\end{tabular}
\end{table}

This faint AGN may drive thermal fluctuations of the IGM through  $\HeII$ photoheating \citep{Bolton2012}.  Although outside the DEIMOS wavelength coverage, $\NV$ emission indicates that there should be photons above 54.4 eV to ionise $\HeII\rightarrow\HeIII$. Using the EUV spectral slope of $\alpha_Q=1.5$, the $\HeII$ ionising photon production is $\dot{N}_{\rm ion}^{\HeII}\approx6.6\times10^{54}\rm~photons~s^{-1}$. The size of the $\HeIII$ region so produced is
\begin{align}
R_{\rm HeIII}&=\left[\frac{3}{4\pi}\frac{\dot{N}_{\rm ion}^{\HeII} t_Q}{(Y/X)\bar{n}_{\rm H}(z)}\right]^{1/3}, \nonumber \\
&\approx3.2\left(\frac{\dot{N}_{\rm ion}^{\HeII}}{6.6\times10^{54}\rm~s^{-1}}\right)^{1/3}\left(\frac{t_Q}{10^8\rm~yr}\right)^{1/3}\rm~pMpc.
\end{align}
A fiducial AGN lifetime of order $10^8\rm~yr$ can be estimated from the timescale required to grow the relevant SMBH. For Eddington-limit accretion, even a massive $100\rm~M_\odot$ black hole seed requires $t=t_{\rm BH}\ln(M_{\rm BH}/M_{\rm seed})\approx5.8\times10^8\rm~yr$ where $t_{\rm BH}\approx4.4\times10^7(\epsilon_r/0.1)\rm~yr$. Therefore, based on the BH growth timescale and outflow timescale\footnote{The upper age limit can be estimated from the non-detection of metal line absorbers at the redshift of RD J1148+5253 AGN. There is broad absorption blueward of $\NV$ $\lambda1240$, indicating an outflow of $v_{\rm outflow}\approx1000-10000\rm~km~s^{-1}$. This constrains the AGN lifetime to $t\lesssim d_{\rm spike}/v_{\rm outlow}\approx3\times10^{8-9}\rm~yr$ where $d_{\rm spike}\approx 3.1\rm~pMpc$.} arguments, during the plausible AGN lifetime, the two broad Ly$\alpha$ transmission spikes lie within the region of influence of the $\HeIII$ I-front of the AGN. 

The $\HeII$ photoheating across the $\HeIII$ I-front raises the temperature approximately by \citep[e.g.][]{Kakiichi2017}
\begin{equation}
\Delta T_{\rm HeIII}=\frac{2}{3k_{\rm B}}\frac{G_{\rm HeII}/\Gamma_{\rm HeII}}{2(X/Y)+3}\approx 6400\left(\frac{2+\alpha_Q^{\rm eff}}{2.5}\right)^{-1},
\end{equation}
where we adopt a EUV spectral index of $\alpha^{\rm eff}_Q=0.5$ to include the effect of spectral hardening $\alpha_Q^{\rm eff}<\alpha_Q$, resulting in the IGM temperature of $\simeq16000\rm~K$ immediately after the $\HeIII$ I-front \citep{Meiksin2010,Bolton2012,Ciardi2012,Khrykin2017}. Such $\HeII$ photoheating reduces the optical depth by a factor of $(1.6\times10^4{\rm~K}/10^4{\rm~K})^{-0.72}=0.71$. While a small decrease, now a slightly less underdense gas below $\Delta_b\lesssim0.30$ can give rise to a transmission spike. This $\HeII$ heating by AGN doubles the occurrence probability of a transmission spike from $P(<\Delta_b^{\rm th}=0.25)=3.6~\%$ at $T=10^4\rm~K$ to $P(<\Delta_b^{\rm th}=0.30)=8.0~\%$ at $T=1.6\times10^4\rm~K$. The AGN $\HeII$ photoheating is also a convenient hypothesis as the spatially coherent increase in the Ly$\alpha$ transmission in the $\HeII$ photoheated region could produce broader transmission spikes (e.g $220\rm~km~s^{-1}$ corresponds to $\simeq330\rm~pkpc$ patch of the IGM) whereas the other spikes widths ($100\rm~km~s^{-1}$ are of order the Jeans length of the $\HI$-photoionized IGM. 

Hence, the association between the $z\simeq5.7$ AGN and the proximate broad Ly$\alpha$ transmission spikes suggests that while faint AGN are unlikely a main driver of $\HI$ reionisation, the hard ionising spectra of AGN may be important to drive the spatial fluctuations of the ionisation and thermal state of the IGM, via possibly an early onset of $\HeII$ reionisation.

\section{Discussion and summary}\label{sec:discussion}

We have initiated a spectroscopic programme involving the 3-D mapping of $5<z<7$ galaxies around the Ly$\alpha$ forest region illuminated by background QSOs which enables us to examine the ionising capabilities of galaxies and AGN at high redshift. In this paper we describe a science verification of this method using DEIMOS spectroscopy of $5.3<z<6.4$ LBGs in the SDSS J1148+5251 field.

Although our sample of confirmed sources is modest, cross-correlation of the spectroscopically-confirmed LBGs with the Ly$\alpha$ forest, reveals tentative, but promising, evidence for a ``statistical $\HI$ proximity effect" indicating that the Ly$\alpha$ transmission of the IGM is preferentially higher in the vicinity of the galaxies. We have interpreted this signal as evidence for an enhanced UV background around luminous LBGs caused by their ionising radiation together with that arising from fainter undetected sources clustered around them. We demonstrate that the required ionising radiation from the luminous LBGs alone is insufficient. This conclusion is supported by independence evidence from deeper imaging observations as well as the expectations of hierarchical clustering in $\Lambda$CDM cosmology. This explanation for the statistical $\HI$ proximity effect is preferred over alternative hypotheses based solely on gas density or thermal fluctuations of the IGM. Such explanations would produce an anti-correlation yielding an excess Ly$\alpha$ absorption around galaxies. Only UV background fluctuations driven by ionising radiation from galaxies can predict the $\HI$ proximity effect.
Balancing the UV background required by the statistical $\HI$ proximity effect with the abundance of spectroscopically-confirmed LBGs and their fainter associates has enabled us to constrain the {\it average escape fraction of LyC photons} at $\langle f_{\rm esc}\rangle\simeq$ $0.08^{+0.08}_{-0.02}$ with $M_{\rm UV}^{\rm lim}\simeq-15\pm3$ at $z\simeq5.8$ using the CLF/HOD framework. 

The present method for constraining $f_{\rm esc}$ has some advantages over previous approaches. It examines the direct influence of galaxies on the local IGM  as well as the bias of ionising sources estimated from  the galaxy-Ly$\alpha$ cross-correlation; this allows us to deduce the relative contributions of luminous and feeble sources as well as that of AGN. The largest uncertainty at present arises from application to a single QSO sightline and small number statistics. Fortunately,
this is easy to remedy with further observations. While a number of assumptions have been made in deriving this value of $f_{\rm esc}$, we have argued that the uncertainties affecting assumed values for the mean free path and thermal fluctuations in the IGM are likely to increase the derived fraction, strengthening the conclusion that the galaxy population is capable of driving cosmic reionisation. Fundamental to our method however, is the assumption that our spectroscopically-confirmed sample is unbiased and independent of the surrounding gaseous environment. Since the bulk of our redshifts are based on detecting Ly$\alpha$ emission, if such photons are attenuated by nearby gas this may lower the spectroscopic success rate and may bias the cross-correlation. Such  a problem may however be mitigated by examining Ly$\alpha$ haloes as the postulated reduced visibility of Ly$\alpha$ line from galaxies would still produce a bright halo detectable with integral field spectroscopy \citep{Kakiichi2017b}.

As discussed above, the widely-held view that the abundant population of intrinsically faint galaxies drives cosmic reionisation is supported by this work. This is also consistent with the belief that the typical escape fraction rises at higher redshift as younger, lower-mass, galaxies are more susceptible to feedback from intense star-forming activity creating a porous interstellar medium \citep[e.g.][]{Kimm2014,Wise2014}. Although there is evidence that reionisation may be accelerated around luminous star-forming galaxies \citep{Stark2017}, the statistical $\HI$ proximity effect can only be understood if there are intrinsically fainter galaxies clustered around the luminous systems. We note this need not conflict with suggestions that the some of the most luminous systems have harder ionising radiation \citep{Laporte2017}.

Finally, we explored the specific role of one luminous LBG and a faint AGN where proximate transmission spikes can be directly (as opposed to statistically) associated. A discovery of individual transverse proximity effect via a Ly$\beta$ transmission spike in the vicinity of a luminous LBG at $z$=6.177 suggests that luminous star-forming systems preferentially reside in highly ionised environments. This supports a deduction from the high fraction of Ly$\alpha$ emission in luminous LBGs at $z$>6 \citep{Curtis-Lake2012,Stark2017}, for which the visibility of Ly$\alpha$ is boosted by large ionised bubbles \citep[e.g.][for a review]{2016ASSL..423..145D}. Accelerated reionisation around the luminous system likely requires clustered fainter galaxies, whose presence may be indicated by excess $\OI$ absobers \citep{Becker2006}. This scenario may gain further support from an observed galaxy overdensity around a pair of bright Ly$\alpha$ emitting galaxies at $z\sim7$ \citep{2011ApJ...730L..35V,Castellano2016}. The broad Ly$\alpha$ transmission spikes in the vicinity of a $z$=5.701 faint AGN suggests that the hard ionising spectra may have an important contribution to the large-scale spatial fluctuations of the UV background and thermal state of the IGM. An interesting possibility is that a patchy early ($z$>5.7) onset of $\HeII$ reionisation by AGN \citep{Bolton2012} heats the IGM  through $\HeII$ photoionisation heating. This late-time $\HeII$ heating induces thermal fluctuations so that the intergalactic Ly$\alpha$ opacity is preferentially reduced around luminous systems, without conflicting with the observed statistical $\HI$ proximity effect. This may explain the large scatter of intergalactic Ly$\alpha$ opacity at the tail end of reionisation \citep{Becker2015a} without need for a large ($\gtrsim$50\%) contribution of AGN to the UV background (\citealt{Chardin2015,Chardin2017,Chardin2018,DAloisio2017}, see also \citealt{Finlator2016}) or extreme thermal injection via $\HII$ photoheating at the early time of $\HI$ reionisation \citep{DAloisio2015}.

Putting all together, a hypothesis emerging from the initial DEIMOS spectroscopy in the QSO field J1148+5251 is that while the faint galaxies with high escape fraction primarily drive reionization, luminous galaxies and AGN may play an increasingly important role towards the end of the reionization process by sourcing the large-scale spatial fluctuations of the UV background and thermal state of the IGM. This demonstrates the potential of spectroscopic survey of 5<$z$<7 galaxies toward QSO fields for making a progress with existing facilities before the JWST and Extremely Large Telescopes, allowing us to tackle the most challenging aspect of cosmic reionisation: {\it "What reionised the Universe?"}.

\section*{Acknowledgments}

We acknowledge financial support from European Research Council Advanced Grant FP7/669253 (KK, NL, RSE, RM, SB). DPS acknowledges support from the National Science Foundation through grant AST-1410155. ERW acknowledges support from the Australian Research Council Centre of Excellence for All Sky Astrophysics in 3 Dimensions (ASTRO 3D), through project number CE170100013.
It is a pleasure to thank the following for useful discussion: Will Hartley, Andreu Font-Ribera, Harley Katz, Tom Fletcher, Jamie Bolton, Masami Ouchi, and Akio Inoue. We acknowledge useful email correspondence with Fred Davies and Ali Rahmati regarding
their simulations. We thank the referee for useful comments and, in
particular, on suggestions for improving the analysis with larger datasets. The data presented herein were obtained at the W. M. Keck Observatory, which is operated as a scientific partnership among the California Institute of Technology, the University of California and the National Aeronautics and Space Administration. The Observatory was made possible by the generous financial support of the W. M. Keck Foundation. The authors wish to recognize and acknowledge the very significant cultural role and reverence that the summit of Mauna kea has always had within the indigenous Hawaiian community.  We are most fortunate to have the opportunity to conduct observations from this beautiful mountain. This work was undertaken using the UCL Grace High Performance Computing Facility (Grace@UCL) and we thank the associated support services. 
BER acknowledges partial support through NASA contract NNG16PJ25C, and grants 17-ATP17-0034 and HST-GO-14747.

\bibliographystyle{mnras}
\bibliography{Reference}

\appendix

\section{Theoretical framework}\label{app:model}
\subsection{Cosmological radiative transfer}\label{app:RT}

Here we present a more complete treatment of cosmological radiative transfer of ionising photons. The equation of cosmological radiative transfer follows \citep[e.g][]{Gnedin1997,Meiksin2009}
\begin{equation}
\frac{1}{c}\frac{\partial I_\nu}{\partial t}+{\boldsymbol n}\cdot{\boldsymbol \nabla} I_\nu-\frac{H}{c}\left(\nu\frac{\partial I_\nu}{\partial \nu}-3HI_\nu\right)
=-\alpha_\nu I_\nu+\varepsilon_\nu,\label{eq:RT}
\end{equation}
where $\alpha_\nu$ is the absorption coefficient and $\varepsilon_\nu$ is the emissivity. The direct solution to this clearly requires expensive numerical radiative transfer simulations. Instead we seek an approximate statistical solution following the approach of \citet{Zuo1992a,Zuo1992b,Meiksin2003,Kakiichi2012}. First, consider a small patch of the universe at position $\br$ with volume $V$ so that the cosmological redshifting can be ignored, which should have a minor impact at $z>5$ \citep{Becker2013}. The number of galaxies above a luminosity $L_{\rm min}$ in the patch follows the Poisson distribution $P(N)=\bar{N}^Ne^{-\bar{N}}/N!$ with the mean value $\bar{N}$. 

As we are interested in the radiation field around a spectroscopically-detected LBG, we split the specific intensity into the contribution from the central LBG $J_0(\br)$ and surrounding galaxies $J_\nu(\br)$. Then, in a patch with $N$ galaxies around a LBG with specific luminosity $L_{\mbox{\tiny LBG}}$, integrating Equation (\ref{eq:RT}) we find that the specific intensity at a distance $\br$ from the LBG is given by
\begin{equation}
I_\nu(\br)=J_0(\br)+J_\nu(\br),
\end{equation}
where
\begin{align}
J_0(\br)=\frac{L_{\mbox{\tiny LBG}}e^{-\tau_c(|\br|)}}{(4\pi)^2|\br|^2},~~J_\nu(\br)=\sum_{k=1}^N \frac{L_k e^{-\tau_c(|\br-\br_k|)}}{(4\pi)^2|\br-\br_k|^2},\label{eq:A3}
\end{align}
and $L_k$ is the specific luminosity and $\br_k$ is the proper distance of $k$-th galaxy from the LBG, and $\tau_c$ is the optical depth of ionising photons over a distance $|\br-\br_k|$. 

To derive the statistically-averaged specific intensity around LBGs, we take the ensemble-averaging over many realisations of patches with various numbers of galaxies. Using the statistical method of characteristic functions \citep{Meiksin2003,Kakiichi2012} or otherwise \citep{Zuo1992a,Zuo1992b}, this gives the average specific intensity, 
\begin{equation}
\langle I_\nu(\br)\rangle=\langle J_0(\br)\rangle+\sum_{N=0}^\infty P(N)\int J_\nu(\br) P[J_\nu(\br)|N]dJ_\nu(\br),\label{eq:A4}
\end{equation}
where $P[J_\nu(\br)|N]$ is the probability distribution function of specific intensity in a patch with $N$ galaxies. When the positions and luminosities of surrounding galaxies are statistically independent to each other (but can be correlated with the LBG)  (e.g. van Kampen 2007), we may express $P[J_\nu(\br)|N]$ as a product of the probablities of finding each galaxy at a position $\br_k$ with a luminosity $L_k$,
\begin{align}
&P[J_\nu(\br)|N]dJ_\nu(\br)=\prod_{k=1}^N \frac{\Phi(L_k)dL_k}{\bar{n}_g(>L_{\mbox{\tiny min}})}\left[1+\xi_g(\br_k,L_k)\right]\frac{d^3r_k}{V},\label{eq:A5}
\end{align}
where $\bar{n}_g(>L_{\mbox{\tiny min}})=\int_{L_{\mbox{\tiny min}}}^\infty \Phi(L)dL$ and $\xi_g(\br,L)$ is the correlation function of LBGs with galaxies of luminosity $L$. Therefore, by substituting Equations (\ref{eq:A3}) and (\ref{eq:A5}) into (\ref{eq:A4}) we obtain, after some algebra,
\begin{equation}
\langle I_\nu(\br)\rangle
=\langle J_0(\br)\rangle+\bar{\varepsilon}_\nu\int \frac{\langle e^{-\tau_c(|\br-\br'|)}\rangle}{(4\pi)^2|\br-\br'|^2}\left[1+\langle\xi_g(\br')\rangle_L\right]d^3r',\label{eq:A6}
\end{equation}
where the local contribution from the LBG is given by $\langle J_0(\br)\rangle=\langle L_{\mbox{\tiny LBG}}\rangle \langle e^{-\tau_c(|\br|)}\rangle/(4\pi|\br|)^2$ and $\bar{\varepsilon}_\nu=\int_{L_{\mbox{\tiny min}}}^\infty L\Phi(L)dL$ is the mean emissivity of galaxies. Expressions such as Equation (\ref{eq:A6}) are intuitive and quoted elsewhere in literature \citep[e.g.][]{Mas-Ribas2016}. The above statistical derivation elucidates various assumptions and clarifies that the average specific intensity depends on the {\it luminosity-weighted correlation function} defined as
\begin{equation}
\langle\xi_g(\br)\rangle_L=\frac{\int^\infty_{L_{\rm min}} L\Phi(L)\xi_g(\br,L)dL}{\int^\infty_{L_{\rm min}} L\Phi(L)dL}.
\end{equation}

In the derivation, because we are dealing with the propagation of photons in an ensemble-averaged sense, we have replaced the attenuation along an individual sightline $e^{-\tau_c(|\br-\br_k|)}$ with the average value $\langle e^{-\tau_c(|\br-\br_k|)}\rangle$ and approximated as
\begin{equation}
\langle e^{-\tau_c(|\br-\br'|)}\rangle\approx e^{-|\br-\br'|/\lambda_{\rm mfp}}.
\end{equation}
The mean free path of ionising photons is given by
\begin{equation}
\lambda_{\rm mfp}^{-1}=\left|\frac{dz}{dl_p}\right|\int d\NHI\CDDF\left[1-e^{-\sigmaHI(\nu)\NHI }\right],
\end{equation}
for Poisson-distributed absorbers \citep{Paresce1980,HM2012}. The mean free path depends on the number density of $\HI$ absorbers, which is quantified by the $\HI$ column density distribution function (CDDF) $\CDDF$. Parametrising the CDDF as a power-law $\CDDF\propto\NHI^{-\beta_{\mbox{\tiny N}}}$ (e.g. $\beta_{\mbox{\tiny N}}=1.33\pm0.05$, \citealt{Becker2013}), it may be written as $\lambda_{\rm mfp}=\lambda_{912}(\nu/\nu_{912})^{3(\beta_{\mbox{\tiny N}}-1)}$ \citep{FG2008} where $\lambda_{912}$ is the mean free path of Lyman-limit photons. In this paper, we assume a constant mean free path at Lyman-limit $\lambda_{\rm mfp}\approx\lambda_{912}$. This produces a systematic error, underestimating the mean photoionisation rate by a small factor of $[3+\alpha_g-3(\beta_{\mbox{\tiny N}}-1)]/(3+\alpha_g)\approx0.84$ when $\alpha_g=3$ and $\beta_{\mbox{\tiny N}}=1.33$ because ignoring the effect that higher frequency photons can reach longer distance before being attenuated. Although adopting a constant spatially uniform mean free path is clearly a oversimplification, it gives a first-order approximation to the mean free path. To encapsulate this model uncertainty (see Section \ref{sec:physical_processes}), we use the Gaussian prior on $\lambda_{\rm mfp}$ with variance of $2\rm~pMpc$.

Furthermore, Equation (\ref{eq:A6}) can be written more succinctly in Fourier space. By realising that Equation (\ref{eq:A6}) is the convolution between the radiative transfer kernel $\frac{e^{-|\br-\br'|/\lambda_{\rm mfp}}}{(4\pi\lambda_{\rm mfp})|\br-\br'|^2}$ and the luminosity-depedent correlation function $\langle \xi_g(\br)\rangle_L$, we arrive at 
\begin{align}
&\langle I_\nu(r)\rangle= \langle J_0(r)\rangle+\nonumber \\
&~~~~~~\frac{\bar{\varepsilon}_\nu\lambda_{\rm mfp}}{4\pi }\left[1+\int_0^\infty \frac{k^2dk}{2\pi^2} R(k\lambda_{\rm mfp})\langle P_g(k)\rangle_L\frac{\sin kr}{kr}\right],\label{eq:A10}
\end{align}
where $R(x)=\arctan(x)/x$ comes from the Fourier transform of the radiative transfer kernel and the luminosity-dependent galaxy power spectrum is
\begin{equation}
\langle P_g(k)\rangle_L=\int_0^\infty \langle \xi_g(r)\rangle_L 4\pi r^2\frac{\sin kr}{kr}dr.
\end{equation}
The expression reduces to the local approximation of the Poisson-distributed sources $\langle J_\nu(r)\rangle=\bar{\varepsilon}_\nu\lambda_{\rm mfp}/(4\pi)$ \citep[e.g.][]{Schirber2003} when there is no galaxy clustering around LBGs, $\langle P_g(k)\rangle_L=0$. 

Finally, we suppose that all galaxies have the same spectral energy distribution with the EUV (>13.6 eV) slope $\alpha_{\rm g}$ to evaluate a typical photoionisation rate at a distance $r$ from a LBG, in which the EUV emissivity from star-forming galaxies is 
\begin{equation}
\bar{\varepsilon}_\nu=h\alpha_g\left(\frac{\nu}{\nu_{912}}\right)^{-\alpha_g}\dot{n}_{\rm ion}(>L_{\rm min}).
\end{equation}
Hence, using the approximate statistical solution (\ref{eq:A10}) of the radiation field, we obtain the typical photoionisation rate at a distance $r$ from a LBG:
\begin{align}
&\langle{\Gamma}_{\rm HI}(r)\rangle=\int_{\nu_{\rm HI}}^\infty\sigma_{\rm HI}(\nu)\frac{4\pi \langle{I}_\nu(r)\rangle}{h\nu}d\nu, \nonumber \\
&=\langle\Gamma_{\rm HI}^{\mbox{\tiny LBG}}(r)\rangle+\bar{\Gamma}_{\rm HI}\left[1+\int_0^\infty \frac{k^2dk}{2\pi^2} R(k\lambda_{\rm mfp})\langle P_g(k)\rangle_L\frac{\sin kr}{kr}\right],\label{eq:A13}
\end{align}
where the first term $\langle\Gamma_{\rm HI}^{\mbox{\tiny LBG}}(r)\rangle=\frac{\alpha_{\rm g} \sigma_{\mbox{\tiny 912}}}{\alpha_{\rm g}+3}\frac{\langle\dot{N}_{\rm ion}^{\mbox{\tiny LBG}}\rangle}{4\pi r^2}e^{-r/\lambda_{\rm mfp}}$ is the local contribution from the central LBGs and the second term is the clustering contribution from the surrounding galaxies.  We use Equation (\ref{eq:A13}) throughout the analysis presented in this paper.

\subsection{Galaxy abundance from HOD framework}\label{app:HOD}

\begin{figure*}
\centering
\includegraphics[width=\textwidth]{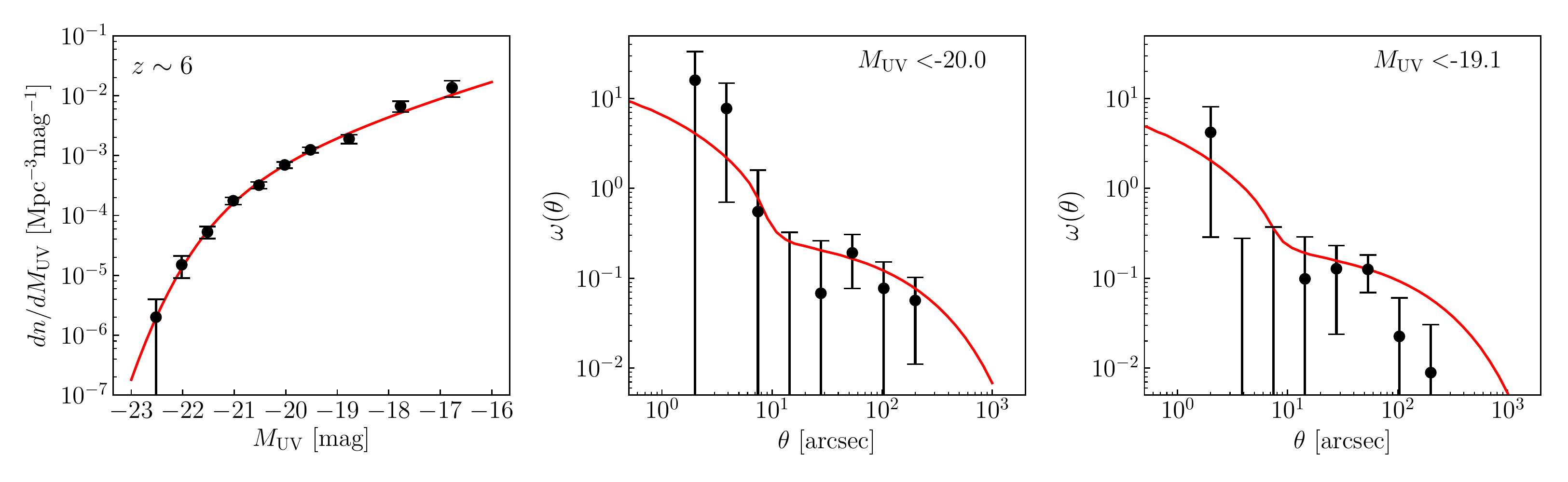}
\vspace{-0.5cm}
\caption{Comparison of the best-fit CLF model with the $z\sim6$ UV luminosity function of Bouwens et al (2015) and LBG angular correlation functions of Harikane et al (2016) (red: model, black: observations).}\label{fig:CLF}
\end{figure*}

We use the HOD framework to estimate the number of fainter, undetected, galaxies clustered around LBGs. We use the conditional luminosity function (CLF) approach \citep[e.g.][]{Yang2003} to the halo occupation distribution (HOD) framework. The CLF, $\Phi(L|M_h)$, specifies the average number of galaxies with luminosities in the range of $L\pm dL/2$ that reside in a halo of mass $M_h$. Thus, by combining the best-fit CLF with the theoretical estimate of the clustering of dark matter haloes around the LBG-host haloes from $N$-body simulations \citep[e.g.][]{Tinker2008,Tinker2010}, we can infer the number of (undetected) galaxies around the (observed) LBGs. To this end, we constrain the CLF model by simultaneously fitting to the observed UV luminosity function \citep{Bouwens2015} and angular correlation function of $z\sim6$ LBGs \citep{Harikane2016}. We follow the CLF model of \citet{vandenBosch2013}, which is summarised below. We drop the subscript $_{\rm UV}$ of $L_{\rm UV}$ for notational clarity.

We split the CLF from the contribution from central galaxies, $\Phi_{\rm cen}(L|M_h)$, and satellite galaxies,  $\Phi_{\rm sat}(L|M_h)$:
\begin{equation}
\Phi(L|M_h)=\Phi_{\rm cen}(L|M_h)+\Phi_{\rm sat}(L|M_h).
\end{equation}
We model the CLF of central galaxies model as a log-normal distribution,
\begin{equation}
\Phi_{\rm cen}(L|M_h)dL=\frac{\log_{10}e}{\sqrt{2\pi}\sigma_{\rm c}}\exp\left[-\frac{(\log_{10}L-\log_{10}L_{\rm c})^2}{2\sigma_{\rm c}^2}\right]\frac{dL}{L},
\end{equation}
where $\sigma_c$ quantifies the scatter in UV luminosity of central galaxies and halo mass and
we adopt a following parameterisation for the central UV luminosity - halo mass relation,
\begin{equation}
L_{\rm c}(M_h)=L_0\frac{(M_h/M_h^\ast)^{\gamma_1}}{[1+(M_h/M_h^\ast)]^{\gamma_1-\gamma_2}},
\end{equation}
where $L_0$ is the normalization, $M_h^\ast$ is a characteristic halo mass, $\gamma_1$ and $\gamma_2$ are the power-law slope at low-mass ($M_h\ll M_h^\ast$) and high-mass ($M_h\gg M_h^\ast$) ends, respectively. The CLF for satellite galaxies is modelled as a modified Schechter function,
\begin{equation}
\Phi_{\rm sat}(L|M_h)dL=\phi_s^\ast\left(\frac{L}{L_s^\ast}\right)^{\alpha_s+1}\exp\left[-\left(\frac{L}{L_s^\ast}\right)^2\right]\frac{dL}{L},
\end{equation}
where $L_s^\ast(M_h)=0.562L_{\rm c}(M_h)$ \citep{Yang2008} and
\begin{equation}
\phi_s^\ast(M_h)=\phi_0\left(\frac{M_h}{10^{12}h^{-1}{\rm M_\odot}}\right)^{\beta_s}.
\end{equation}
Therefore, the CLF model contains eight free parameters, $\boldsymbol{\theta}_{\rm CLF}=(\log L_0,\log M_h^\ast,\gamma_1,\gamma_2,\sigma_c, \log \phi_0,\alpha_s,\beta_s)$ (strictly speaking, we express  $\log L_0$ in terms of the corresponding UV magnitude $M_{\rm UV,0}$). 

Once the CLF is specified, we can compute the luminosity function and the correlation function (power spectrum) of galaxies. The luminosity function is given by
\begin{equation}
\Phi(L)=\int \Phi(L|M_h)\frac{dn}{dM_h}dM_h,
\end{equation}
where $dn/dM_h$ is the halo mass function for which we use \citet{Tinker2008} halo mass function.

The galaxy power spectrum is computed using the standard HOD framework. Using the CLF, the halo occupation number of central galaxies above a limiting luminosity threshold of sample $L_{\rm th}$ is given by,
\begin{equation}
\langle N_{\rm cen}|M_h\rangle=\int_{L_{\rm th}}^\infty \Phi_{\rm cen}(L|M_h)dL, 
\end{equation}
and for satellite galaxies, $\langle N_{\rm sat}|M_h\rangle=\int_{L_{\rm th}}^\infty \Phi_{\rm sat}(L|M_h)dL$. The number density of galaxies is $\bar{n}_g(>L_{\rm th})=\int \langle N|M_h\rangle \frac{dn}{dM_h}dM_h$ where $\langle N|M_h\rangle=\langle N_{\rm cen}|M_h\rangle+\langle N_{\rm sat}|M_h\rangle$ is the total halo occupation number of galaxies. In the halo model \citep[e.g.][]{Cooray2006}, the power spectrum of galaxies is expressed in terms of one-halo and two-halo terms containing all possible combinations of central and satellites,
\begin{align}
P_g(k)=&2P^{1h}_{{\rm cs}}(k)+P^{1h}_{{\rm ss}}(k) \nonumber \\
&+P^{2h}_{{\rm cc}}(k)+2P^{2h}_{{\rm cs}}(k)+P^{2h}_{{\rm ss}}(k).
\end{align}
Following the notation of \citet{vandenBosch2013}, we have defined the necessary one-halo $P^{1h}_{\rm {xy}}(k)$ and two-halo terms $P^{2h}_{\rm {xy}}(k)$ as
\begin{equation}
P^{1h}_{\rm {xy}}(k)=\int dM_h \mathcal{H}_{\rm x}(k,M_h)\mathcal{H}_{\rm y}(k,M_h) \frac{dn}{dM_h},
\end{equation}
and 
\begin{align}
P^{2h}_{\rm {xy}}(k)=P_m(k)&\int dM_h \mathcal{H}_{\rm x}(k|M_h) b_h(M_h) \frac{dn}{dM_h} \nonumber \\
 &\times \int dM'_h \mathcal{H}_{\rm y}(k|M'_h) b_h(M'_h) \frac{dn}{dM'_h}
\end{align}
where `x' and `y' are either 'c' (for central) or `s' (for satellite), and
\begin{equation}
\mathcal{H}_c(k,M_h)=\frac{\langle N_{\rm cen}|M_h\rangle}{\bar{n}_g(>L_{\rm th})},~~
\mathcal{H}_s(k,M_h)=\frac{\langle N_{\rm sat}|M_h\rangle}{\bar{n}_g(>L_{\rm th})}\tilde{u}(k|M_h).\label{eq:Hx}
\end{equation}
For the dark matter power spectrum $P_m(k)$, we use the non-linear fitting formula of \citet{Peacock1996}. The result is  marginally affected even if we use the linear matter power spectrum. For the halo bias factor $b_h(M_h)$, we adopt the fitting function of \citet{Tinker2010}. Here, $\tilde{u}(k|M_h)$ is the Fourier transform of the NFW halo profile and for the halo concentration parameter we use \citet{Duffy2008} fitting function.

Finally, we compute the angular correlation function of galaxies from the galaxy power spectrum. Using the Limber approximation, the angular correlation function at a perpendicular separation $r_\perp$ is given by
\begin{equation}
\omega_{ij}(r_\perp)=\int dz N^2(z)\left| \frac{d\chi}{dz}\right|^{-1}\int \frac{dk}{2\pi}kP_{ij}(k)J_0(kr_\perp),
\end{equation}
where $N(z)$ is the normalized redshift distribution of galaxies, $\left|d\chi/dz\right|=c/H(z)$, and $J_0(kr_\perp)$ is the zeroth-order Bessel function of the first kind. We use $N(z)$ from the Monte Carlo simulation of $i$-dropouts by \citet{Bouwens2015}.

\begin{table}
\centering
\caption{The best-fit CLF parameters}\label{table:CLF_parameters}
\begin{tabular}{ll}
\hline\hline
parameter &  best-fit value \\
\hline
$M_{\rm UV,0}$ & $-21.43^{+1.36}_{-0.96}$ \\
$\log M_h^\ast$ & $11.56^{+0.28}_{-0.43}$ \\
$\gamma_1$ & $2.10^{+0.55}_{-0.25}$ \\
$\gamma_2$ & $0.25^{+0.36}_{-0.48}$ \\
$\sigma_c$ & $0.2$ (fixed) \\
$\log \phi_0$ & $-0.94^{+0.08}_{-0.04}$ \\
$\alpha_s$ & $-1.15^{+0.11}_{-0.18}$ \\
$\beta_s$ & $1.11^{+0.45}_{-0.53}$ \\
\hline
\end{tabular}
\end{table}

To specify the CLF parameters, we simultaneously fit the model with the $z\sim6$ UV luminosity function of \citet{Bouwens2015} and the angular correlation function of LBGs of \citet{Harikane2016}. using the Markov chain Monte Carlo (MCMC) method using \textsc{emcee} package \citep{Foreman-Mackey2013}. We assume a Gaussian likelihood and only use the diagonal element of the error covariance matrix. We assume flat priors for all CLF parameters. The best-fit parameters are computed as the 50 percentiles of the posterior distributions. We use the best-fit CLF parameters in the analysis throughout this paper. The result of joint fitting procedure is shown in Figure \ref{fig:CLF} and the best-fit parameters are tabulated in Table \ref{table:CLF_parameters}. With a larger dataset we can readily improve our analysis by simultaneously fitting the CLF parameters, $\langle f_{\rm esc}\rangle$, and $M_{\rm UV}^{\rm lim}$ in a full MCMC framework. For the purpose of the paper, we keep the CLF parameters to be fixed at the best-fit values for simplicity.

For the application to cosmological radiative transfer, we need to specify the luminosity-dependent cross-power spectrum between our LBG samples ($M_{\rm UV}<-21$) and galaxies with luminosity $L$. In halo model, this is given by
\begin{align}
P_g(k,L)=&P^{1h}_{\rm cs}(k,L)+P^{1h}_{\rm sc}(k,L)+P^{1h}_{\rm ss}(k,L) \nonumber \\
&+P^{2h}_{\rm cc}(k,L)+P^{2h}_{\rm cs}(k,L)+P^{2h}_{\rm sc}(k,L)+P^{2h}_{\rm ss}(k,L).\label{eq:PkL}
\end{align}
where 
\begin{equation}
P^{1h}_{\rm {xy}}(k,L)=\int dM_h \mathcal{H}_{\rm x}(k,M_h)\mathcal{C}_{\rm y}(k,L,M_h) \frac{dn}{dM_h},
\end{equation}
\begin{align}
P^{2h}_{\rm {xy}}(k,L)=&P_m(k)\int dM_h \mathcal{H}_{\rm x}(k,M_h) b_h(M_h) \frac{dn}{dM_h} \nonumber \\
 &~~~~~~~~~\times \int dM'_h \mathcal{C}_{\rm y}(k,L,M'_h) b_h(M'_h) \frac{dn}{dM'_h},
\end{align}
and $\mathcal{H}_{\rm x}(k,M_h)$ is defined in the same way as Equation (\ref{eq:Hx}) but using a luminosity threshold $L_{\rm th}$ corresponding to our LBG samples, and
\begin{equation}
\mathcal{C}_c(k,L,M)=\frac{\Phi_{\rm cen}(L|M_h)}{\Phi(L)},~
\mathcal{C}_s(k,M)=\frac{\Phi_{\rm sat}(L|M_h)}{\Phi(L)}\tilde{u}(k|M_h).
\end{equation}
Finally, using the best-fit CLF parameters we evaluate and substitute Equation (\ref{eq:PkL}) into the luminosity-weighted galaxy power spectrum, Equation (\ref{eq:14}), to model the enhanced photoionisation rate around LBGs throughout this paper.

\section{Linear theory}\label{app:B}

\begin{figure}
\centering
\includegraphics[width=\columnwidth]{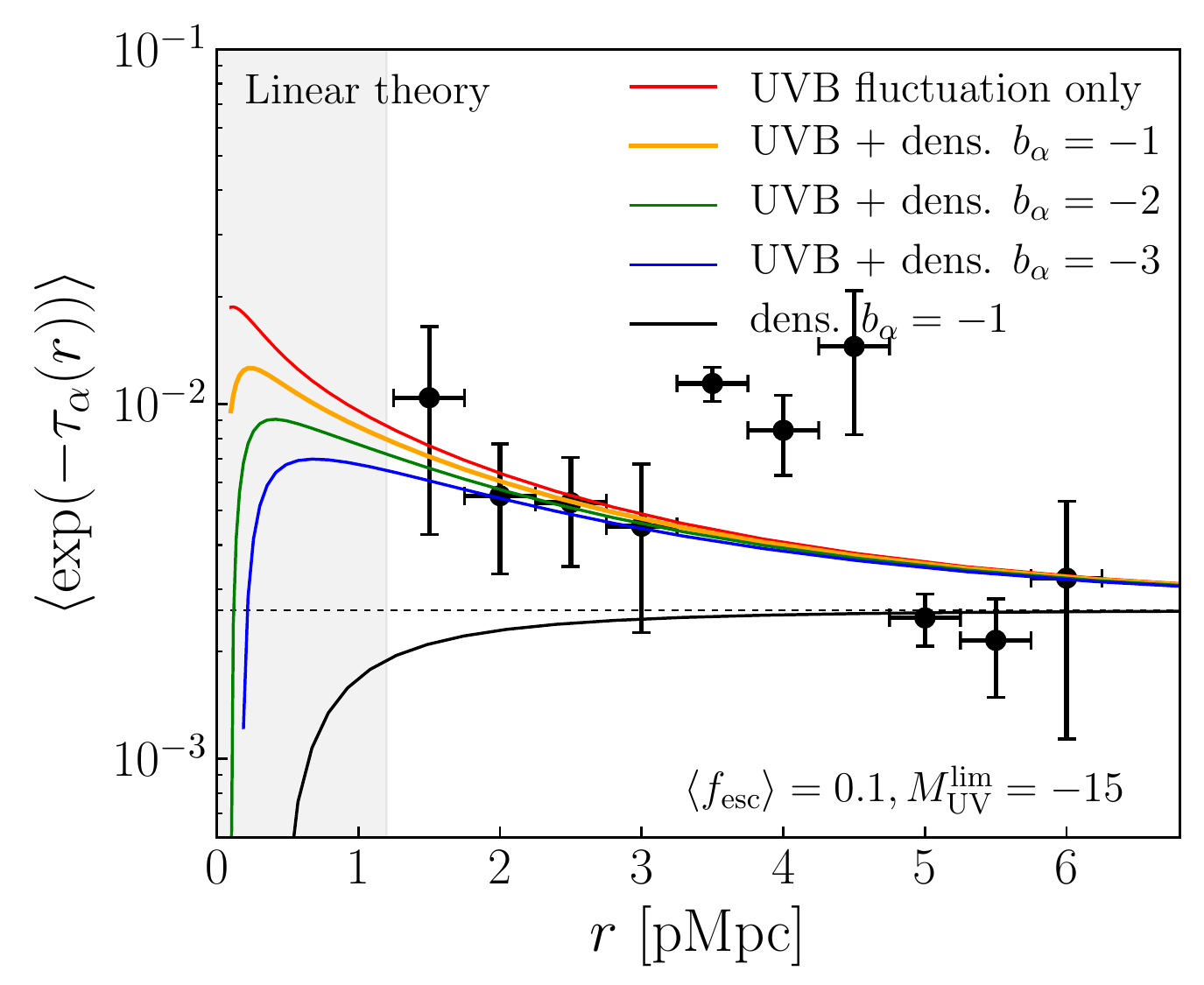}
\vspace{-0.5cm}
\caption{Effect of the galaxy-gas density correlation on the mean Ly$\alpha$ transmitted flux around LBGs in the linear regime. The BOSS-based estimate of Ly$\alpha$ forest bias $b_\alpha\simeq -1$ at $z=5.8$ (orange) shows only a modest effect of matter correlation on the galaxy-Ly$\alpha$ forest cross-correlation on the large-scale ($\gtrsim$1 pMpc) presented in the paper.  The galaxy-gas density correlation only case is also shown (black). All models assume $\langle f_{\rm esc}\rangle=0.1$, $M_{\rm UV}^{\rm lim}=-15$, $\lambda_{\rm mfp}=6$ pMpc, $T=10^4$ K, and the best-fit CLF parameters.}\label{fig:linear}
\end{figure}

Here we quantify the effect of galaxy-gas density correlation on the mean Ly$\alpha$ transmitted flux around LBGs. While in the main analysis we use a fully nonlinear galaxy-galaxy correlation function in the UV background around LBGs $\langle \Gamma_{\rm HI}(r)\rangle$ (Appendix~\ref{app:model}), to examine the relative contribution of galaxy-galaxy and galaxy-gas density correlations, we use the linear theory so that a fair comparison of the two competing effects can be made at the same linear order. 

Taylor expanding our model of the mean Ly$\alpha$ transmitted flux around LBGs (Equation (\ref{eq:model})) in terms of the photoionisation rate, we find
\begin{equation}
\langle \exp(-\tau_\alpha(r))\rangle\approx \bar{F}_\alpha \left[1+b_\Gamma\langle\delta_\Gamma(r)\rangle\right], 
\end{equation}
where $\bar{F}_\alpha=\int d\Delta_b P_V(\Delta_b) e^{-\bar{\tau}_\alpha(\bar{\Gamma}_{\rm HI},T)\Delta_b^2}$ is the mean Ly$\alpha$ transtmitted flux of the IGM. The UV background fluctuation $\langle \delta_\Gamma(r)\rangle=\langle \Gamma_{\rm HI}(r) \rangle/\bar{\Gamma}_{\rm HI}-1$ reduces to
\begin{equation}
\langle \delta_\Gamma(r)\rangle=b_{\mbox{\tiny LBG}}\langle b_g\rangle_L\int\frac{k^2 dk}{2\pi^2}R(k\lambda_{\rm mfp})P_m^{\rm lin}(k)\frac{\sin kr}{kr},
\end{equation}
in the linear regime, and the bias factor is the response of the Ly$\alpha$ transmitted flux to a small perturbation of UV background,
\begin{align}
b_\Gamma&=\frac{1}{\bar{F}_\alpha}\left.\frac{d\langle F_\alpha\rangle}{d\langle \delta_\Gamma\rangle}\right|_{\langle \delta_\Gamma\rangle=0}, \nonumber \\
&=\frac{1}{\bar{F}_\alpha}\int d\Delta_b P_V(\Delta_b)\bar{\tau}_\alpha(\bar{\Gamma}_{\rm HI},T)\Delta_b^2 
e^{-\bar{\tau}_\alpha(\bar{\Gamma}_{\rm HI},T)\Delta_b^2}.
\end{align}
This shows that our nonlinear model is equivalent to the well-known linear theory \citep{Font-Ribera2013,duMasdesBourboux2017} at the correct limit. 

Thus, following the linear theory model, the contribution of galaxy-gas density correlation can be included as \citep{Font-Ribera2013,duMasdesBourboux2017}
\begin{equation}
\langle \exp(-\tau_\alpha(r))\rangle\approx \bar{F}_\alpha \left[1+b_\Gamma\langle\delta_\Gamma(r)\rangle+b_{\mbox{\tiny LBG}}b_\alpha \xi_m^{\rm lin}(r)\right], 
\end{equation}
where $b_\alpha$ is the Ly$\alpha$ forest bias factor and $\xi_m^{\rm lin}(r)$ is the linear matter correlation function. We estimate the Ly$\alpha$ forest bias using the BOSS Ly$\alpha$ forest result $b_\alpha(z)\simeq-0.134[(1+z)/(1+2.4)]^{2.9}$ \citep{Slosar2011,duMasdesBourboux2017}, leading $b_\alpha\simeq-1$ at $z=5.8$. To complement this large extrapolation, we also examine the cases with  $b_\alpha\simeq-2$ and $-3$.

In Figure~\ref{fig:linear} we show the effect of the galaxy-gas density correlation on the mean Ly$\alpha$ transmitted flux around LBGs. The increasing mean gas overdensity around LBGs reduces the Ly$\alpha$ transmission at smaller radii as argued in the main text. The effect would become prominent only at smaller scale ($\lesssim$1 pMpc), which is below the scale presented in the paper. The relative contribution is below 10 per cent for the BOSS-based estimate $b_\alpha\simeq-1$ at the innermost bin (1.5 pMpc), and only modestly increases with Ly$\alpha$ forest bias at the scale of interest. The effect of galaxy-gas density correlation should thus be small. Note that, regardless of the precise value of the effect, the contribution of galaxy-gas density correlation requires {\it more} ionising photons to match the observed Ly$\alpha$ transmitted flux in order to compensate the mean gas overdensity, leading to an even higher value of escape fraction. Our main result will therefore remain unchanged.

\label{lastpage}

\end{document}